\documentclass[12pt,letterpaper]{article}
\usepackage[a4paper, total={7in, 10in}]{geometry}

\usepackage{graphicx}
\usepackage{helvet}
\usepackage{authblk}
\usepackage{hyperref}
\usepackage{amsmath} 
\usepackage{amssymb} 
\usepackage{orcidlink} 
\usepackage[super,comma,sort&compress]  
   {natbib}
\usepackage[right]{lineno} \linenumbers
\usepackage[title]{appendix}%

\usepackage[utf8]{inputenc} 
\usepackage[T1]{fontenc}    
\usepackage{url}            
\usepackage{booktabs}       
\usepackage{amsfonts}       
\usepackage{nicefrac}       
\usepackage{microtype}      
\usepackage{xcolor}         
\usepackage{multirow}
\usepackage{threeparttable}
\usepackage{amsmath}
\usepackage{subcaption}
\usepackage{makecell}
\usepackage{placeins}
\usepackage[utf8]{inputenc} 
\usepackage{booktabs}      
\usepackage{siunitx}       
\usepackage{caption}       
\usepackage{tabularx}
\usepackage{listings}    
\usepackage{pifont}
\usepackage{array}
\usepackage{makecell}
\theadset{\bfseries}

\renewcommand{\textcolor}[2]{#2}

\makeatletter
\renewcommand{\maketitle}{\bgroup\setlength{\parindent}{0pt}
\begin{flushleft}
  \textbf{\@title}
  
  \@author
\end{flushleft}\egroup}
\makeatother

\title{Artificial Intelligence-derived Photoplethysmography Age as a Digital Biomarker for Cardiovascular Health}
\date{}

\author[1,2,$\#$]{Guangkun Nie}
\author[3,$\#$]{Qinghao Zhao}
\author[1,2]{Gongzheng Tang}
\author[1]{Yaxin Li}
\author[1,2,4,5,*]{Shenda Hong}

\affil[1]{Institute of Medical Technology, Health Science Center of Peking University, Beijing, China}
\affil[2]{National Institute of Health Data Science, Peking University, Beijing, China}
\affil[3]{Department of Cardiology, Peking University People’s Hospital, Beijing, China}
\affil[4]{State Key Laboratory of Vascular Homeostasis and Remodeling, NHC Key Laboratory of Cardiovascular Molecular Biology and Regulatory Peptides, Peking University, Beijing, China}
\affil[5]{Department of Emergency Medicine, Peking University First Hospital, Beijing, China}
\affil[$\#$]{These authors contributed equally}
\affil[*]{Correspondence: hongshenda@pku.edu.cn}

\begin{document}

\maketitle

\section*{ABSTRACT}
\textbf{Background:} Photoplethysmography (PPG), increasingly available through wearable devices, provides a non-invasive means of monitoring human hemodynamics. In this study, we introduce artificial intelligence-derived photoplethysmography (AI-PPG) age, a deep learning-based estimate of biological age from raw PPG signals, and evaluate its potential as a digital biomarker for cardiovascular health.

\textbf{Methods:} We developed a deep learning model with a distribution-aware loss function to reduce bias from imbalanced data. The model was trained and evaluated on the UK Biobank cohort (N = 212,231). We analyzed the association between the AI-PPG age gap (AI-PPG age minus calendar age) and multiple cardiovascular and metabolic outcomes, assessed its longitudinal value using serial PPG measurements, and externally validated its generalizability in an independent MIMIC-III-derived cohort (N = 2,343).

\textbf{Results:} After adjusting for key confounders, participants with an AI-PPG age gap greater than 9 years have a significantly higher risk of major adverse cardiovascular and cerebrovascular events (hazard ratio of 2.37, \textcolor{blue}{p = 8.46$\times$10$^{-80}$}), as well as seven secondary outcomes including coronary heart disease and myocardial infarction (all p < 0.005). Conversely, those with a gap below -9 years show a lower risk profile. Longitudinal analysis demonstrates that changes in AI-PPG age add predictive value over time. In the external validation cohort, each one-year increase in AI-PPG age gap is associated with higher in-hospital mortality (odds ratio of 1.02, p = 0.01).

\textbf{Conclusions:} AI-PPG age is a scalable, non-invasive biomarker for cardiovascular health assessment. Integrated with wearable devices, it may enable population-level screening, personalized monitoring, and early intervention.

\section*{Plain language summary}
Wearable devices can measure tiny changes in blood flow using light. We developed a computer method that turns this information into a measure called "PPG age", which shows how old the blood vessels appear. In our study of over 200,000 people, those with a PPG age much higher than their actual age were more likely to develop heart problems, such as coronary heart disease. People with a younger PPG age had lower risks. Tracking changes in this measure over time also provided useful clues about future health. Because it works with simple wearable sensors, this approach could support large-scale heart health screening and personalized prevention in everyday life.

\section*{INTRODUCTION}
Cardiovascular diseases (CVDs) are the leading cause of global mortality and a major contributor to reduced quality of life \cite{roth2018global, cooper2018global}. In 2021, an estimated 621 million individuals worldwide were affected by CVDs, resulting in approximately 20.5 million deaths \cite{lindstrom2022global}. Early risk identification and timely intervention are crucial for improving outcomes and preventing disease progression \cite{van2024risk, wong2024pan}. However, existing screening and diagnostic methods primarily depend on specialized healthcare professionals and facility-based equipment \cite{vogel2021lancet, pandey2025systematic, addissouky2024recent, li2023artificial}, which limits accessibility, particularly in resource-constrained settings. Moreover, the lack of simple, user-friendly tools for cardiovascular risk assessment outside clinical environments hinders timely monitoring and proactive health management. This gap affects not only individuals already diagnosed with CVDs but also those at elevated risk who could benefit from earlier intervention. These challenges highlight the pressing need for accessible, cost-effective solutions that provide clear, actionable insights into cardiovascular health, enabling individuals to conveniently monitor their risk and promoting the broader adoption of preventive care strategies.

Recent advancements in sensor technology and data analytics have significantly broadened the potential applications of photoplethysmography (PPG) signals \cite{nie2024review}. PPG is a non-invasive and easily accessible optical physiological signal that can effectively monitor hemodynamic parameters, such as heart rate \cite{pankaj2022review} and blood oxygen saturation \cite{schroder2023apple}, and has been widely integrated into wearable devices including smartwatches \cite{pereira2020photoplethysmography, williams2023wearable, krittanawong2021integration}. With the growth of artificial intelligence (AI), the application of PPG in CVDs, such as atrial fibrillation \cite{yang2019using, pereira2020photoplethysmography} and hypertension \cite{elgendi2019use, mousavi2019blood}, has gained increasing attention. Studies have shown that vascular aging, which leads to increased arterial stiffness, is closely associated with a heightened risk of CVDs \cite{vasan2019interrelations, vasan2022arterial} and induces characteristic changes in PPG waveforms \cite{li2022xgboost, chen2024predicting}. Given this relationship, it is conceivable to extract quantitative biomarkers from PPG signals that can assess the extent of vascular aging and be used for cardiovascular health evaluation and management.

Building on the success of AI-driven digital biomarkers such as AI-ECG age \cite{lima2021deep, park2024artificial, hempel2025explainable} and AI-brain age \cite{chen2025accelerometer}, and recognizing the intrinsic relationship between vascular aging and PPG waveform characteristics, we hypothesize that AI can directly estimate a biological age from raw PPG signals (AI-PPG age), serving as a marker of vascular aging and an indicator of vascular health. In healthy individuals, since vascular properties generally correspond to their calendar age, the AI-PPG age closely aligns with the calendar age. By contrast, individuals with CVDs tend to exhibit an AI-PPG age that exceeds their calendar age, as their vascular properties resemble those of older individuals, causing their PPG waveforms to similarly reflect advanced vascular aging. This rationale supports the use of a deep learning model trained to predict calendar age from PPG signals in healthy individuals as a tool for assessing vascular aging and as a non-invasive method for evaluating vascular health.

In this study, we present a deep learning-based approach to extract AI-PPG age. To address the challenge of imbalanced age distribution in the training data, we introduce a distribution-aware loss function to mitigate this problem. The model is developed and evaluated using the UK Biobank (UKB) dataset. Our results show that the AI-PPG age gap, defined as the difference between AI-PPG age and calendar age, serves as an independent risk factor for the primary outcome of major adverse cardiovascular events (MACCE) as well as for secondary outcomes, including diabetes, hypertension, coronary heart disease (CHD), heart failure (HF), myocardial infarction (MI), stroke, and all-cause mortality. Additionally, we assess the longitudinal utility of AI-PPG age by analyzing serial PPG data from the same individuals over time. External validation on a MIMIC-III-derived dataset further confirms that the AI-PPG age gap is positively associated with in-hospital mortality. These findings highlight the potential of AI-PPG age as a valuable tool for cardiovascular health monitoring and risk management.

\begin{figure}
    \centering
    \includegraphics[width=0.96\linewidth]{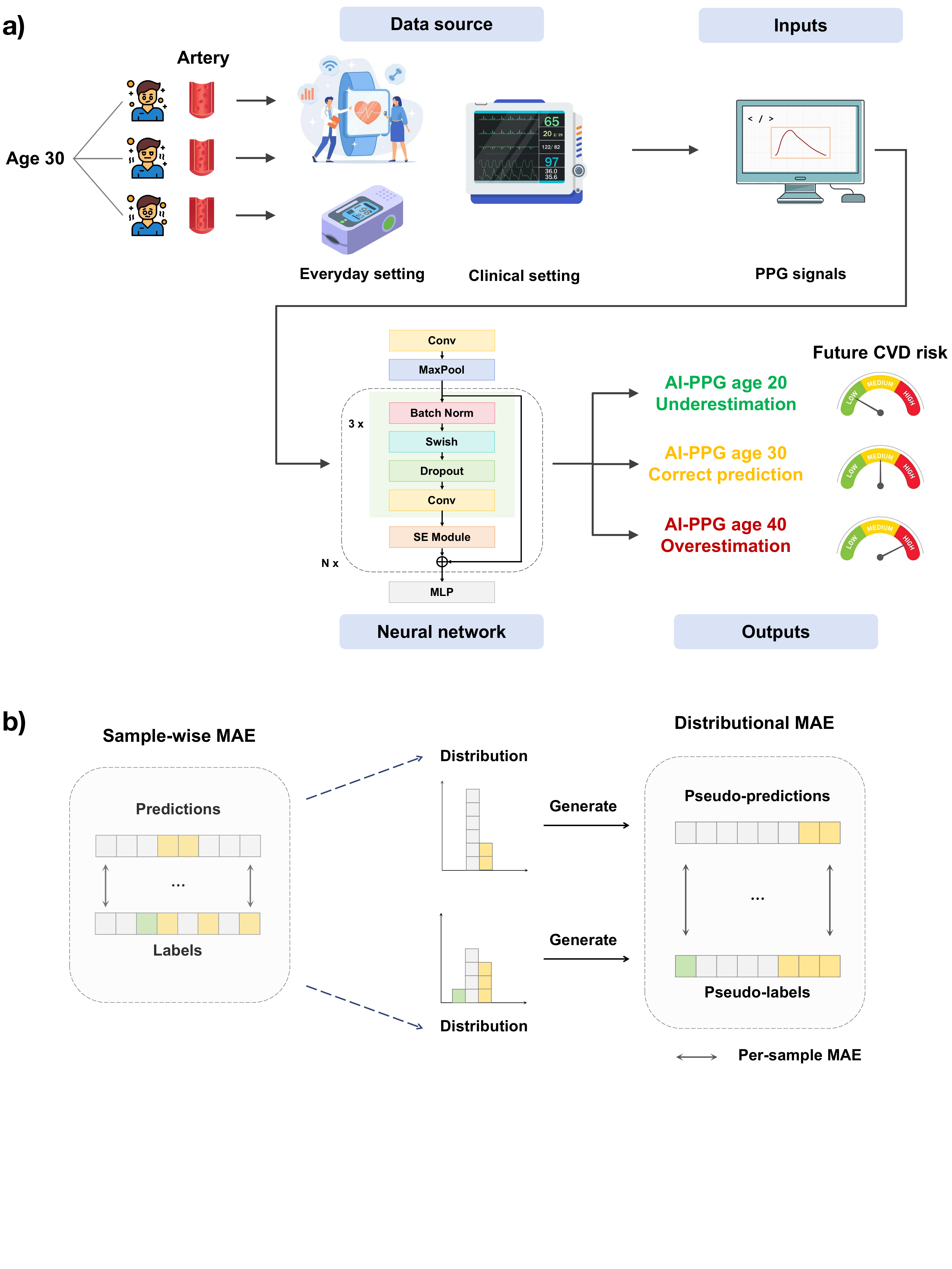}
    \caption{Development and application of the AI-PPG age prediction model. a) We employ a deep learning model to predict AI-PPG age from raw PPG waveforms. A higher AI-PPG age gap, defined as the difference between AI-PPG age and calendar age, is associated with an increased risk of future CVDs. b) In the training phase, we introduce Dist Loss to address the imbalanced regression problem. Dist Loss comprises two components: sample-wise MAE and distributional MAE. The sample-wise MAE quantifies the discrepancy between predictions and ground truth labels, while the distributional MAE captures the divergence between their respective distributions. To illustrate the core concept behind distributional MAE, the figure simplifies its computation by assuming that the batch size equals the total number of training samples. CVDs, cardiovascular diseases; MACCE, major adverse cardiovascular and cerebrovascular events; MAE, mean absolute error.}
    \label{fig: framework}
\end{figure}

\section*{Methods}
\subsection*{Datasets}
\paragraph{UK Biobank cohort}
The UKB is a large-scale biomedical resource comprising de-identified genetic, lifestyle, and health data, along with biological samples, from over 500,000 participants across the United Kingdom \cite{sudlow2015uk}. The cohort includes four assessment instances: the initial assessment visit (2006-2010) with 502,150 participants, the first repeat assessment visit (2012-2013) with 20,337 participants, the imaging visit (2014-present) with 90,683 participants, and the first repeat imaging visit (2019-present) with 12,734 participants. Longitudinal follow-up data are also available, enabling comprehensive tracking of participants' health over time. In this study, PPG signals were extracted from UKB field 4205, and corresponding age information was obtained from UKB field 21003. The PPG signals were recorded using the PulseTrace PCA2 device, with each signal preprocessed and provided by UKB in a standardized format containing 100 sample points per waveform. 

\paragraph{PulseDB Dataset}
PulseDB is a curated dataset designed specifically for benchmarking cuff-less blood pressure estimation methods \cite{wang2023pulsedb}. It contains high-quality 10-second PPG segments sampled at 125 Hz from 5,361 subjects, accompanied by demographic information such as age. The dataset consists of two primary subsets: the MIMIC-III subset (n=2,423) and the VitalDB subset (n=2,938). The MIMIC-III subset includes PPG data from patients admitted to the critical care units at Beth Israel Deaconess Medical Center between 2001 and 2012 \cite{johnson2016mimic_iii}, while the VitalDB subset comprises PPG recordings from surgical patients undergoing routine or emergency procedures at Seoul National University Hospital, Republic of Korea \cite{lee2022vitaldb}. In this study, the VitalDB subset was used to fine-tune the AI-PPG age prediction model initially developed on the UKB dataset. This fine-tuning step was necessary to account for potential temporal and population-level differences between the UKB and PulseDB datasets, ensuring better generalization and adaptability of the model to new data sources, while the MIMIC-III subset was employed for external validation. Furthermore, the MIMIC-III subset within PulseDB was linked to the MIMIC-III Clinical Database \cite{johnson2016mimic} to obtain corresponding in-hospital mortality data for additional analysis.

Given the high quality of the PPG data in both the UKB and PulseDB datasets, minimal preprocessing was required. Z-score normalization was applied to standardize the signals and ensure consistency across subjects, aligning input data distributions for robust model training.

\subsection*{Model development}
\subsubsection*{Model architecture}
The AI-PPG age prediction model is based on Net1D \cite{hong2020holmes}, a one-dimensional convolutional neural network (CNN) built on a ResNet \cite{he2016deep} variant. It is selected for its ability to capture subtle and discriminative patterns in PPG signals related to vascular aging. Net1D employs sequential convolutional blocks with residual connections \cite{he2016deep} and squeeze-and-excitation (SE) modules \cite{hu2018squeeze} to enhance feature extraction. The residual connections mitigate model degradation and stabilize training, while the SE modules adaptively recalibrate channel-wise feature responses to emphasize relevant vascular aging features. This architecture ensures robust and efficient processing of PPG signals for AI-PPG age prediction. In our implementation, the Net1D-based model comprises 389,689 learnable parameters, and the detailed model configuration and pretrained weights are provided (\url{https://huggingface.co/Ngks03/PPG-VascularAge}).

\subsubsection*{Imbalanced regression in AI-PPG age prediction}
In AI-PPG age prediction using the UKB dataset, a significant challenge arises from the highly imbalanced age distribution. Most samples are concentrated within the age range of 50 (second decile) to 69 (ninth decile) years, while the tail regions (37-49 and 70-87 years) contain relatively few samples. Conventional loss functions, such as the mean absolute error (MAE), treat all samples equally. Consequently, models trained with these loss functions tend to focus excessively on the majority age group. This causes samples in sparsely populated regions to be biased toward the dense central region, resulting in a pronounced deviation between the distribution of model predictions and the true label distribution. To address this issue, we propose Dist Loss, a distribution-aware loss function that enhances performance in underrepresented regions without sacrificing overall accuracy. The central idea of Dist Loss is to align the distribution of model predictions with the label distribution, thereby ensuring accurate predictions across all age groups (Figure \ref{fig: framework}).

The technical details of the Dist Loss are described below, and further pseudocode is available in Supplementary Figure S2. Dist Loss comprises two main components: the sample-wise MAE and the distributional MAE. The sample-wise MAE is equivalent to the conventional MAE loss and computes the absolute error between each prediction and its corresponding ground truth, ensuring individual sample accuracy. The distributional MAE, on the other hand, quantifies the discrepancy between the distribution of model predictions and the true label distribution. It is optimized by minimizing the MAE between a pseudo-label sequence (constructed based on the label distribution) and a pseudo-prediction sequence (obtained by sorting the model predictions).

Specifically, we first employ kernel density estimation (KDE)~\cite{parzen1962estimation} to obtain a smooth approximation of the label distribution across the dataset. For each label \( y_i \), a probability \( p_i \) is estimated to reflect its overall frequency. For a batch of size \( B \), the expected frequency of \( y_i \) is given by \( n_i = B \cdot p_i \). Since \( n_i \) is generally non-integer, we take its floor value \( \lfloor n_i \rfloor \). To ensure that the total frequency matches the batch size \( B \), we compute the residual \( r = B - \sum_{i=1}^L \lfloor n_i \rfloor \), where \( L \) denotes the number of unique labels. An auxiliary sequence \( \mathcal{R} = (r_1, r_2, \dots, r_L) \) is then defined as 
\[
r_i = 
\begin{cases} 
1, & \text{if } i \le \left\lfloor \frac{r+1}{2} \right\rfloor \text{ or } i > B - \left\lfloor \frac{r}{2} \right\rfloor, \\
0, & \text{otherwise}, \tag{1}
\end{cases}
\]
which distributes the residual across the labels. The adjusted frequency for each label is given by \( n'_i = \lfloor n_i \rfloor + r_i \), ensuring that \( \sum_{i=1}^L n'_i = B \). Using these adjusted frequencies, we construct the pseudo-label sequence \( \mathcal{S}_L = (S_{L_1}, S_{L_2}, \dots, S_{L_B}) \) that reflects the theoretical label distribution. For each position \( j \) in the batch (\( 1 \leq j \leq B \)), the pseudo-label is determined by 
\[
S_{L_j} = y_{\, \arg\min_{1 \le i \le L} \left( \sum_{k=1}^i n_k \geq j \right)}, \tag{2}
\]

where \( j = 1, \dots, B \) indexes the position in the sequence. Essentially, this procedure replicates each label \( y_i \) according to its adjusted frequency \( n'_i \). For instance, if the label set is \( (40, 50, 60) \) with adjusted frequencies \( (1, 2, 3) \), the resulting pseudo-label sequence is \( (40, 50, 50, 60, 60, 60) \). The pseudo-prediction sequence \( \mathcal{S}_P \) is obtained by sorting the model predictions in the batch. To ensure differentiability, we employ the fast differentiable sorting method described in~\cite{blondel2020fast}.

The overall loss function is then defined as
\[
L_{\text{Dist}} = \mathrm{MAE}(\mathbf{Y}, \hat{\mathbf{Y}}) + \mathrm{MAE}(\mathcal{S}_L, \mathcal{S}_P), \tag{3}
\]
where \( \mathbf{Y} \) and \( \hat{\mathbf{Y}} \) denote the ground truth labels and the model predictions, respectively. The first term ensures the accuracy of individual sample predictions, while the second term minimizes the discrepancy between the distribution of predictions and the true label distribution. This formulation mitigates the tendency of traditional regression models to overly focus on the majority region, thereby reducing the bias that causes samples in low-sample regions to be predicted toward the dense central region.

In this context, the pseudo-labels and pseudo-predictions can be understood as effective representations of the corresponding distributions (Figure \ref{fig: framework}). To illustrate this, imagine plotting histograms for both the pseudo-labels and pseudo-predictions, where the horizontal axis represents the labels and the vertical axis indicates their frequency. These histograms would reflect the underlying distribution of the labels and predictions. Therefore, the proposed method serves as an approximation of the true label and prediction distributions.

\subsubsection*{Experimental setup}
The model was implemented using the PyTorch 2.3.1 framework and trained on an NVIDIA GeForce RTX 4090 GPU (24GB memory). The training process spanned 80 epochs, utilizing the Adam optimizer with an initial learning rate of 0.003. To enhance training stability, we employed a large batch size of 2,048 and applied L2 weight regularization with a factor of $\lambda = 1 \times 10^{-4}$ to mitigate overfitting. For the Dist loss calculation, Gaussian KDE was used with a bandwidth parameter of $\sigma = 0.5$ to model the age label distribution, which spans 21 to 111 years. Model performance was tracked using the MAE as the primary evaluation metric throughout the training.

\subsubsection*{Model interpretability}
To enhance the interpretability of the model, this study employs saliency maps to visualize the regions of the PPG signals that the model prioritizes when predicting AI-PPG age. Saliency maps are generated by computing the gradients of the input signal with respect to the model’s output, thereby highlighting the model's sensitivity to various components of the signal during prediction. Specifically, regions exhibiting larger gradient magnitudes are indicative of areas that exert a stronger influence on the model’s prediction. To refine the visual clarity, Gaussian smoothing and interpolation techniques are applied to the saliency maps, minimizing noise and improving the discernibility of relevant signal features. 

\subsection*{Outcome definition}
\label{Outcome definition}
This study incorporates several outcome definitions from the UKB dataset, including the primary outcome MACCE and secondary outcomes such as mortality, HF, MI, stroke, CHD, hypertension, and diabetes. MACCE is a composite endpoint comprising mortality, HF, MI, and stroke. Mortality data were obtained from the National Death Registries of the United Kingdom, while MI and stroke outcomes were derived from algorithmically defined criteria. Diagnoses for HF, CHD, hypertension, and diabetes were based on first occurrences. Specifically, CHD outcomes include angina pectoris, acute MI, subsequent MI, complications following acute MI, other acute ischemic heart diseases, and chronic ischemic heart disease. Hypertension encompasses both essential and secondary forms. Diabetes is categorized into insulin-dependent, non-insulin-dependent, malnutrition-related, other specified types, and unspecified diabetes mellitus.

\subsection*{Serial PPG analysis}
Given that some participants in the UKB dataset attended multi-time point assessments, resulting in serial PPG data across different time points, we further investigated the application of AI-PPG age in such serial PPG data. Following the exclusion criteria outlined in Figure \ref{fig: dataflow}, all serial PPG data were excluded from the model development set, leaving 58,232 PPG records from 27,108 participants. Our analysis specifically focused on two distinct time points of serial PPG data: for participants with more than two PPG records, we selected their first two PPG data for analysis. In this analysis, we divided the participants into four groups based on the AI-PPG age gap at the two time points: G1 (overestimation at both time points), G2 (underestimation or correct prediction at the first time point and overestimation at the second), G3 (overestimation at the first time point and underestimation or correct prediction at the second), and G4 (underestimation or correct prediction at both time points). We studied the differences in the risk of future incident MACCE across these groups.

\subsection*{Statistics and Reproducibility}
Statistical and computational analyses were performed using R (version 4.2.2) and Python (version 3.11). In R, the packages \textit{survival} (version 3.8.3), \textit{survminer} (version 0.5.0), \textit{ggplot2} (version 3.5.1), and \textit{rms} (version 7.0.0) were used for survival analyses, visualization, and regression modeling. In Python, libraries including NumPy (v1.26.4), Pandas (v2.2.2), PyTorch (v2.3.1), Lifelines (v0.30.0), and SciPy (v1.14.0) were used for PPG data processing and analysis; visualization in Python used Matplotlib (v3.9.1). Pearson correlation coefficients and MAE were calculated to evaluate the consistency between AI-PPG age and calendar age. The AI-PPG age gap (AI-PPG age minus calendar age) was used as the primary variable for clinical analyses. Multivariable-adjusted Cox proportional hazards regression was applied to assess associations between the AI-PPG age gap and outcomes in the UKB cohort, with hierarchical adjustment models as described in the Methods. Kaplan-Meier (KM) survival curves were compared using log-rank tests. Logistic regression was applied to the MIMIC-III cohort to assess the association with in-hospital mortality.

{For the UKB analyses, the clinical evaluation cohort consisted of 111,561 participants with 141,745 PPG records after excluding individuals with missing BMI or loss to follow-up. Follow-up was administratively censored at December 31, 2022. For the MIMIC-III-derived cohort, analyses included 2,343 participants; BMI was excluded from adjustment models due to a high proportion of missing data. Statistical tests were two-sided, and p-values < 0.05 were considered statistically significant. The reproducibility of our findings is supported by the use of large, publicly available datasets and by providing open access to the code and model weights.

\subsection*{Ethics Statement}
UKB has obtained ethical approval from the North West-Haydock Research Ethics Committee (REC reference: 21/NW/0157). This study was conducted under UKB application number 90018. Access to the MIMIC-III database was granted via PhysioNet after completion of the Collaborative Institutional Training Initiative program, \textcolor{blue}{with approval by the Institutional Review Boards of Beth Israel Deaconess Medical Center and the Massachusetts Institute of Technology.} The PulseDB dataset consists of de-identified, publicly available data derived from MIMIC-III and VitalDB, \textcolor{blue}{and therefore did not require additional ethical review or informed consent in accordance with national legislation and institutional requirements. No further ethical approval was required for this study beyond these existing approvals and provisions.}

\section*{Results}
\subsection*{Study population}
In this study, we developed and evaluated our model using the UKB dataset, PulseDB, and the MIMIC-III Clinical Database (Table \ref{Tab: baseline characteristics}). The data flow across these datasets is depicted in Figure \ref{fig: dataflow}. The UKB dataset contains 243,355 PPG recordings from 212,231 participants and is divided into a development cohort (98,672 PPGs from 98,672 participants) and a clinical evaluation cohort (144,683 PPGs from 113,559 participants). Exclusion criteria for the development cohort were applied at the participant level to prevent data leakage: 1) participants with multiple assessment visits resulting in multiple PPG recordings, and 2) participants with a history of circulatory system disorders, diabetes, or dyslipidemia at the time of assessment. Circulatory disorders were identified according to UKB category 2409. The first criterion retained serial PPG data for longitudinal analysis of AI-PPG age, whereas the second ensured the integrity of training labels. The development cohort was then partitioned into training, selection, and hold-out sets in an 8:1:1 ratio. The selection set was used for hyperparameter tuning and model selection, while the hold-out set was reserved to assess the model's predictive performance. The clinical evaluation cohort was used to assess the predictive capability of AI-PPG age for future clinical events. A subset with serial PPG recordings (N = 27,108, PPG = 54,216) was extracted to form the serial PPG set for longitudinal analysis.

The PulseDB dataset (N = 5,361, PPG = 5,245,454) comprises the VitalDB and MIMIC subsets. For the VitalDB subset, up to ten PPG recordings were randomly selected for each patient to construct a fine-tuning set (N = 2,938, PPG = 29,380). If a patient had fewer than ten recordings, random duplication was applied to reach ten recordings. This fine-tuning set was split into training, selection, and hold-out sets in an 8:1:1 ratio to adapt the UKB-trained model for temporal variations across datasets. In the MIMIC subset, one PPG recording per patient was selected for AI-PPG age prediction and linked to the MIMIC-III Clinical Database to create the MIMIC hospital mortality set (N = 2,343, PPG = 2,343). Given the distinct data sources of the MIMIC and VitalDB subsets, the entire MIMIC hospital mortality set was reserved for external evaluation of the fine-tuned model.

\begin{figure}
    \centering
    \includegraphics[width=0.98\linewidth]{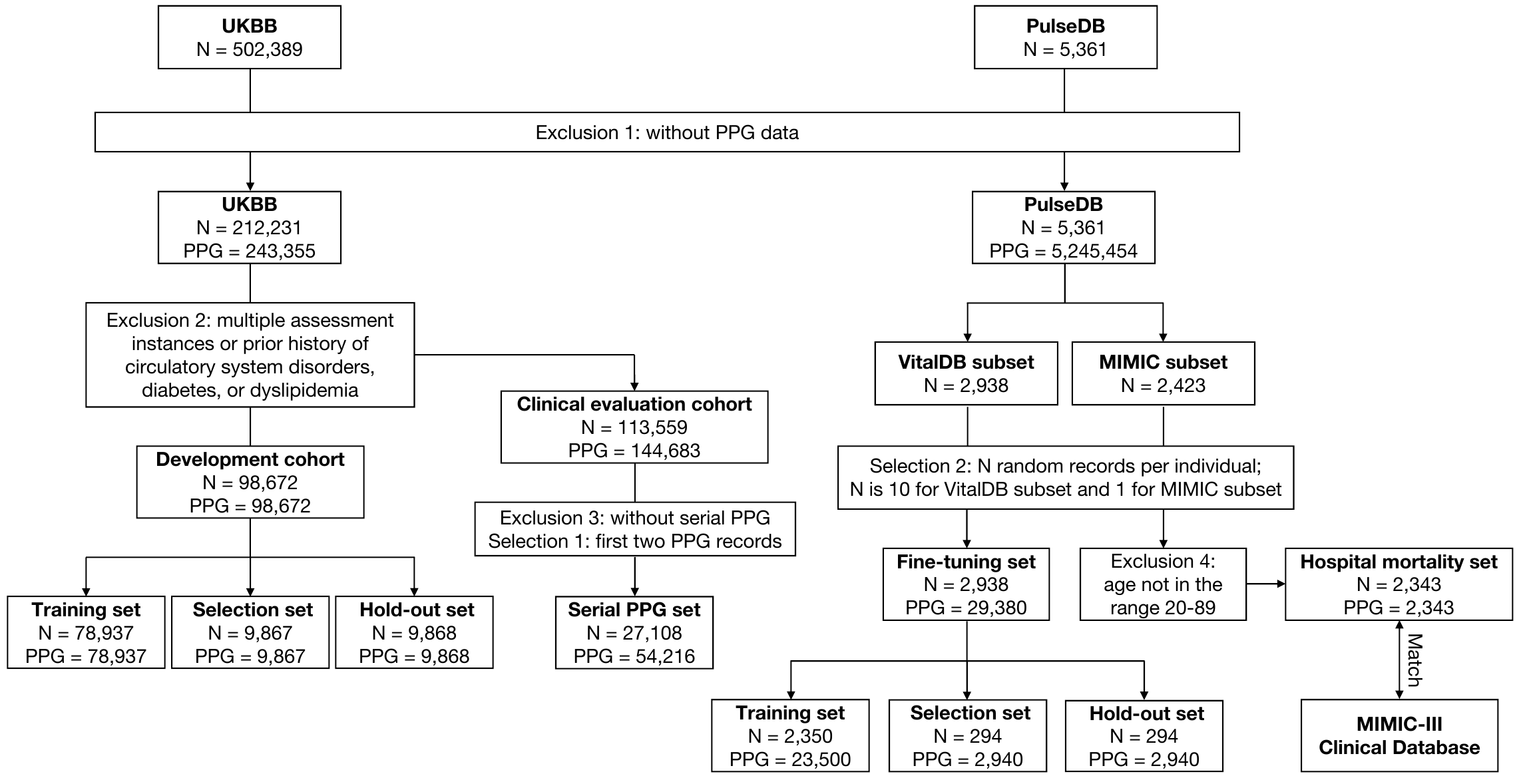}
    \caption{Data flow chart. The AI-PPG age prediction model was developed and evaluated using the UKB dataset. The VitalDB subset of the PulseDB dataset was used to fine-tune the UKB-trained model, while the MIMIC subset of the PulseDB dataset was linked to the MIMIC-III Clinical Database for external evaluation. UKB: UK Biobank.}
    \label{fig: dataflow}
\end{figure}

\begin{table}[]
\begin{threeparttable}
\centering
\caption{Clinical characteristics. UKB, UK Biobank; MACCE: major cardiovascular and cerebrovascular events; NA, not available.}
\label{Tab: baseline characteristics}
\renewcommand{\arraystretch}{1.2}  
\begin{tabular}{lccc}
\hline
Characteristics                     & \multicolumn{2}{c}{UKB}                                                                                                                                    & MIMIC subset                                                               \\ \cline{2-4} 
                                    & \begin{tabular}[c]{@{}c@{}}Development cohort \\ (N=98,672)\end{tabular} & \begin{tabular}[c]{@{}c@{}}Clinical evaluation cohort\\ (N=144,683)\end{tabular} & \begin{tabular}[c]{@{}c@{}}Hospital mortality set\\ (N=2,343)\end{tabular} \\ \hline
Age (years)                         & 55.81±8.63                                                               & 60.87±8.03                                                                       & 61.51±15.25                                                                \\
Sex                                 & N=98,672                                                                 & N=144,683                                                                        & N=2,343                                                                    \\
\multicolumn{1}{r}{Male}            & 40,797 (41.35\%)                                                         & 73,203 (50.60\%)                                                                 & 1,376 (58.73\%)                                                            \\
\multicolumn{1}{r}{Female}          & 57,875 (58.65\%)                                                         & 71,480 (49.40\%)                                                                 & 967 (41.27\%)                                                              \\
Race                                & N=98,065                                                                 & N=143,959                                                                        & N=2,005                                                                    \\
\multicolumn{1}{r}{White}           & 90,398 (92.18\%)                                                         & 134,717 (93.58\%)                                                                & 1,668 (83.19\%)                                                            \\
\multicolumn{1}{r}{Asian}           & 3,329 (3.39\%)                                                           & 4,344 (3.02\%)                                                                   & 47 (2.34\%)                                                                \\
\multicolumn{1}{r}{Black}           & 2,287 (2.33\%)                                                           & 2,750 (1.91\%)                                                                   & 191 (9.53\%)                                                               \\
\multicolumn{1}{r}{Mixed}           & 829 (0.85\%)                                                             & 792 (0.55\%)                                                                     & 6 (0.30\%)                                                                 \\
\multicolumn{1}{r}{Others}          & 1,222 (1.25\%)                                                           & 1,356 (0.94\%)                                                                   & 93 (4.64\%)                                                                \\
BMI (kg/m\textsuperscript{2})                         & 26.46±4.36                                                               & 27.79±4.92                                                                       & 29.41±9.91                                                                 \\
Hypertension                        & 0 (0\%)                                                                  & 69,682 (48.16\%)                                                                 & 1,383 (59.03\%)                                                            \\
Diabetes                            & 0 (0\%)                                                                  & 14,099 (9.74\%)                                                                  & 673 (28.72\%)                                                              \\
Dyslipidemia                        & 0 (0\%)                                                                  & 45,535 (31.47\%)                                                                 & 805 (34.36\%)                                                              \\
Chronic kidney disease              & 454 (0.46\%) & 3639 (2.52\%) & 362 (15.45\%) \\
Smoking status                      & N=98,083                                                                 & N=143,648                                                                        &                                                                            \\
\multicolumn{1}{r}{None-smoker}     & 88,238 (89.96\%)                                                         & 133608 (93.01\%)                                                                 & 2,073 (88.48\%)                                                             \\
\multicolumn{1}{r}{Current smoking} & 9,845 (10.04\%)                                                          & 10,040 (6.99\%)                                                                  & 270 (11.52\%)                                                              \\
Heart Failure                       & 0 (0\%)                                                                  & 1,517 (1.05\%)                                                                   & 559 (23.86\%)                                                              \\
Myocardial infarction               & 8 (0.01\%)                                                               & 5,745 (3.97\%)                                                                   & 383 (16.35\%)                                                              \\
Stroke                              & 11 (0.01\%)                                                              & 3,440 (2.38\%)                                                                   & 323 (13.79\%)                                                              \\
Coronary heart disease              & 0 (0\%)                                                                  & 14,038 (9.70\%)                                                                  & 662 (28.25\%)                                                              \\
MACCE                               & 19 (0.02\%)                                                              & 17,278 (11.94\%)                                                                 & 1,132 (48.31\%)                                                            \\
Follow-up time (years)              & 11.66 (3.10)                                                             & 9.87 (3.23)                                                                      & NA                                                                         \\ \hline
\end{tabular}           
\begin{tablenotes}
\footnotesize
\item  \textcolor{blue}{Continuous variables (e.g., age and BMI) are presented as mean $\pm$ standard deviation. Categorical variables are presented as count (percentage).}
\end{tablenotes}
\end{threeparttable}
\end{table}

\subsection*{Performance of the deep learning model for AI-PPG age prediction}
Performance of the AI-PPG age prediction model was evaluated on the UKB hold-out set, UKB clinical evaluation set, VitalDB hold-out set, and MIMIC hospital mortality set, with detailed results provided in Table \ref{Tab: Model performance}. In the UKB hold-out set, the Pearson correlation coefficient between AI-PPG age and calendar age was 0.49, with an MAE of 7.57 years. In the clinical evaluation set, the correlation was 0.38, with an MAE of 8.18 years, likely due to the inclusion of individuals with existing CVDs, resulting in greater inconsistency between AI-PPG age and calendar age. For the VitalDB hold-out set, the correlation was 0.54 with an MAE of 9.69 years. In the MIMIC hospital mortality set, the correlation dropped to 0.41, with an MAE of 11.81 years, reflecting the greater heterogeneity and hemodynamic instability in ICU patients, which may have contributed to the reduced performance.

\begin{table}
\caption{Performance of the AI-PPG age prediction model. UKB, UK Biobank; MAE, mean absolute error.}
\centering
\renewcommand{\arraystretch}{1.2}  
\begin{tabular}{lcc}
\hline
Datasets                             & Pearson correlation coefficient & MAE (years) \\ \hline
UKB hold-out set   & 0.49  & 7.57 \\
UKB clinical evaluation set & 0.38 & 8.18 \\
VitalDB hold-out set & 0.54 & 9.60  \\
MIMIC hospital mortality set   & 0.41   & 11.81 \\ \hline
\end{tabular}
\label{Tab: Model performance}
\end{table}

\subsection*{AI-PPG age gap as a predictor of major adverse cardiovascular and cerebrovascular events and cardiometabolic diseases}
We utilized the Cox proportional hazards model to assess whether the AI-PPG age gap serves as an independent risk factor for future MACCE, which was designated as the primary outcome in this study. In addition, we explored its association with several secondary outcomes, including diabetes, hypertension, CHD, HF, MI, stroke, and all-cause mortality. To provide a comprehensive evaluation of the predictive value of the AI-PPG age gap, we analyzed it both as a continuous and categorical variable. Three levels of covariate adjustment were applied: Model 1 adjusted for age, sex, ethnicity, and body mass index (BMI); Model 2 further adjusted for smoking status, hypertension, diabetes, dyslipidemia, and chronic kidney disease (CKD); and Model 3 included variables from the Framingham risk score \cite{pencina2009predicting}, namely age, sex, systolic blood pressure (SBP), use of antihypertensive treatment, smoking status, diabetes, total cholesterol, and HDL cholesterol.

In Model 1, restricted cubic spline curves (Figure \ref{Fig: rcs}) were used to illustrate the relationship between the AI-PPG age gap (as a continuous variable) and the risk of MACCE as well as secondary outcomes. As shown, hazard ratios (HRs) progressively increased with the AI-PPG age gap. Table \ref{Tab: UKB HRs} presents the HRs for each one-year increase in the AI-PPG age gap. Notably, significant associations were observed not only for the primary outcome, MACCE (HR = 1.02, p = 2.43$\times$10$^{-100}$), but also for all secondary outcomes: diabetes (HR = 1.04, p  = 1.00$\times$10$^{-128}$), hypertension (HR = 1.05, p = 1.00$\times$10$^{-128}$), CHD (HR = 1.03, p = 1.00$\times$10$^{-128}$), HF (HR = 1.02, p = 8.62$\times$10$^{-26}$), MI (HR = 1.04, p = 1.26$\times$10$^{-77}$), stroke (HR = 1.03, p = 1.89$\times$10$^{-25}$), and all-cause mortality (HR = 1.02, p = 2.17$\times$10$^{-36}$). The trends observed in Model 2 and Model 3 were consistent, with HRs remaining statistically significant for all outcomes.

For further investigation, the study population was stratified into three groups based on the gap: underestimation (< -9 years), correct prediction (-9 to 9 years), and overestimation (> 9 years). The correct prediction group was used as the reference group. We then evaluated the relative risks of future MACCE and secondary outcomes in both the underestimation and overestimation groups. The results, summarized in Table \ref{Tab: UKB HRs}, \textcolor{blue}{revealed that the underestimation group had a significantly lower risk for MACCE (HR = 0.79, p = 2.48$\times$10$^{-36}$) as well as all secondary outcomes, including diabetes (HR = 0.68, p = 1.58$\times$10$^{-39}$), hypertension (HR = 0.62, p = 1.02$\times$10$^{-114}$), CHD (HR = 0.71, p = 4.36$\times$10$^{-56}$), HF (HR = 0.82, p = 4.32$\times$10$^{-11}$), MI (HR = 0.69, p = 9.38$\times$10$^{-32}$), stroke (HR = 0.78, p = 6.52$\times$10$^{-10}$), and all-cause mortality (HR = 0.86, p = 5.47$\times$10$^{-10}$).} In contrast, \textcolor{blue}{the overestimation group exhibited a significantly higher risk for MACCE (HR = 2.37, p = 8.46$\times$10$^{-80}$) and all secondary outcomes, including diabetes (HR = 2.69, p = 3.14$\times$10$^{-77}$), hypertension (HR = 2.88, p = 1.00$\times$10$^{-128}$), CHD (HR = 2.20, p = 4.62$\times$10$^{-58}$), HF (HR = 2.15, p = 2.28$\times$10$^{-19}$), MI (HR = 2.51, p = 5.16$\times$10$^{-35}$), stroke (HR = 2.55, p = 1.53$\times$10$^{-22}$), and all-cause mortality (HR = 2.51, p= 4.85$\times$10$^{-48}$).} These results were consistent in Model 2 and Model 3, where statistical significance was maintained for all outcomes.

Finally, Kaplan-Meier (KM) curves (Figure \ref{Fig: km}) were employed to illustrate the cumulative incidence of MACCE and secondary outcomes across the three AI-PPG age gap groups. The findings highlight the AI-PPG age gap as a robust predictor of MACCE and underscore its potential utility in cardiovascular risk stratification.

\begin{table}[]
\resizebox{\linewidth}{!}{
\begin{threeparttable}
\caption{Adjusted HRs for future MACCE and cardiometabolic diseases. The AI-PPG age gap is evaluated both as a continuous variable and as a categorical variable. For the categorical analysis, participants are classified into three groups: underestimation, correct prediction, and overestimation, with the correct prediction group serving as the reference. HR, hazard ratio; MACCE, major adverse cardiovascular and cerebrovascular events; BMI, body mass index; HDL, high density lipoprotein; SBP, systolic blood pressure.}
\label{Tab: UKB HRs}
\centering
\renewcommand{\arraystretch}{1.2}  
\begin{tabular}{lcccccc}
\hline
Cardiovascular event & \multicolumn{2}{c}{AI-PPG age gap} & \multicolumn{2}{c}{Underestimation} & \multicolumn{2}{c}{Overestimation} \\ \cline{2-7} 
                     & HR (95\% CI)      & p-value          & HR (95\% CI)      & p-value         & HR (95\% CI)     & p-value         \\ \hline
\multicolumn{7}{l}{Model 1, adjusted for age, sex, ethnicity and BMI}                                                           \\
MACCE                & 1.02 (1.02–1.03)  & 2.43 $\times$ 10$^{-100}$ & 0.79 (0.77–0.82)  & 2.48 $\times$ 10$^{-36}$ & 2.37 (2.17–2.59) & 8.46 $\times$ 10$^{-80}$ \\
Diabetes             & 1.04 (1.04–1.05)  & 1.00 $\times$ 10$^{-128}$ & 0.68 (0.64–0.72)  & 1.58 $\times$ 10$^{-39}$ & 2.69 (2.42–2.99) & 3.14 $\times$ 10$^{-77}$ \\
Hypertension         & 1.05 (1.05–1.05)  & 1.00 $\times$ 10$^{-128}$ & 0.62 (0.59–0.64)  & 1.02 $\times$ 10$^{-114}$ & 2.88 (2.66–3.11) & 1.00 $\times$ 10$^{-128}$ \\
Coronary heart disease & 1.03 (1.03–1.04) & 1.00 $\times$ 10$^{-128}$ & 0.71 (0.68–0.74)  & 4.36 $\times$ 10$^{-56}$ & 2.20 (2.00–2.42) & 4.62 $\times$ 10$^{-58}$ \\
Heart failure        & 1.02 (1.02–1.02)  & 8.62 $\times$ 10$^{-26}$  & 0.82 (0.77–0.87)  & 4.32 $\times$ 10$^{-11}$ & 2.15 (1.82–2.54) & 2.28 $\times$ 10$^{-19}$ \\
Myocardial infarction & 1.04 (1.03–1.04) & 1.26 $\times$ 10$^{-77}$  & 0.69 (0.64–0.73)  & 9.38 $\times$ 10$^{-32}$ & 2.51 (2.17–2.91) & 5.16 $\times$ 10$^{-35}$ \\
Stroke               & 1.03 (1.02–1.03)  & 1.89 $\times$ 10$^{-25}$  & 0.78 (0.72–0.85)  & 6.52 $\times$ 10$^{-10}$ & 2.55 (2.11–3.08) & 1.53 $\times$ 10$^{-22}$ \\
All-cause mortality  & 1.02 (1.02–1.02)  & 2.17 $\times$ 10$^{-36}$  & 0.86 (0.82–0.90)  & 5.47 $\times$ 10$^{-10}$ & 2.51 (2.22–2.84) & 4.85 $\times$ 10$^{-48}$ \\ \hline
\multicolumn{7}{l}{Model 2, adjusted for age, sex, ethnicity, BMI, smoking status, hypertension, diabetes, dyslipidemia, and CKD}    \\
MACCE                & 1.02 (1.02–1.02)  & 3.78 $\times$ 10$^{-79}$ & 0.82 (0.79–0.85)  & 2.36 $\times$ 10$^{-28}$ & 2.24 (2.05–2.45) & 1.25 $\times$ 10$^{-69}$ \\
Coronary heart disease & 1.03 (1.03–1.03) & 1.00 $\times$ 10$^{-128}$ & 0.72 (0.69–0.76)  & 1.70 $\times$ 10$^{-49}$ & 2.13 (1.94–2.35) & 4.57 $\times$ 10$^{-53}$ \\
Heart failure        & 1.02 (1.01–1.02)  & 1.84 $\times$ 10$^{-20}$  & 0.84 (0.79–0.89)  & 7.46 $\times$ 10$^{-09}$ & 2.08 (1.76–2.46) & 1.74 $\times$ 10$^{-17}$ \\
Myocardial infarction & 1.03 (1.03–1.04) & 5.12 $\times$ 10$^{-68}$  & 0.70 (0.66–0.75)  & 1.56 $\times$ 10$^{-27}$ & 2.41 (2.08–2.79) & 1.21 $\times$ 10$^{-31}$ \\
Stroke               & 1.02 (1.02–1.03)  & 3.47 $\times$ 10$^{-21}$  & 0.80 (0.74–0.87)  & 3.86 $\times$ 10$^{-08}$ & 2.41 (2.00–2.91) & 6.65 $\times$ 10$^{-20}$ \\
All-cause mortality  & 1.02 (1.01–1.02)  & 5.91 $\times$ 10$^{-25}$  & 0.89 (0.85–0.94)  & 3.28 $\times$ 10$^{-06}$ & 2.35 (2.07–2.66) & 2.80 $\times$ 10$^{-41}$ \\ \hline
\multicolumn{7}{p{20cm}}{
Model 3, adjusted for age, sex, SBP, use of antihypertensive treatment, smoking status, diabetes, total cholesterol, and HDL cholesterol.}
\\
MACCE                & 1.02 (1.02–1.02)  & 1.47 $\times$ 10$^{-56}$  & 0.84 (0.81–0.88)  & 2.04 $\times$ 10$^{-15}$ & 2.37 (2.14–2.62) & 8.71 $\times$ 10$^{-64}$ \\
Coronary heart disease & 1.03 (1.03–1.04) & 3.07 $\times$ 10$^{-96}$  & 0.74 (0.71–0.78)  & 4.01 $\times$ 10$^{-31}$ & 2.33 (2.10–2.59) & 1.37 $\times$ 10$^{-55}$ \\
Heart failure        & 1.02 (1.01–1.02)  & 3.89 $\times$ 10$^{-18}$  & 0.86 (0.80–0.92)  & 1.08 $\times$ 10$^{-05}$ & 2.41 (2.01–2.89) & 3.52 $\times$ 10$^{-21}$ \\
Myocardial infarction & 1.04 (1.03–1.04) & 3.12 $\times$ 10$^{-52}$  & 0.71 (0.66–0.77)  & 5.64 $\times$ 10$^{-19}$ & 2.62 (2.23–3.08) & 2.44 $\times$ 10$^{-31}$ \\
Stroke               & 1.02 (1.02–1.03)  & 3.95 $\times$ 10$^{-13}$  & 0.85 (0.77–0.93)  & 5.10 $\times$ 10$^{-04}$ & 2.33 (1.87–2.91) & 4.31 $\times$ 10$^{-14}$ \\
All-cause mortality  & 1.01 (1.01–1.02)  & 1.38 $\times$ 10$^{-14}$  & 0.94 (0.89–0.99)  & 2.29 $\times$ 10$^{-02}$ & 2.32 (2.02–2.66) & 7.38 $\times$ 10$^{-33}$ \\ \hline
\end{tabular}
\begin{tablenotes}
\footnotesize
\item  \textcolor{blue}{Results in the table were obtained using Cox proportional hazards regression.}
\end{tablenotes}
\end{threeparttable}}
\end{table}

\begin{figure}
\centering
\begin{minipage}{0.24\textwidth}
    \centering
    \includegraphics[width=\linewidth]{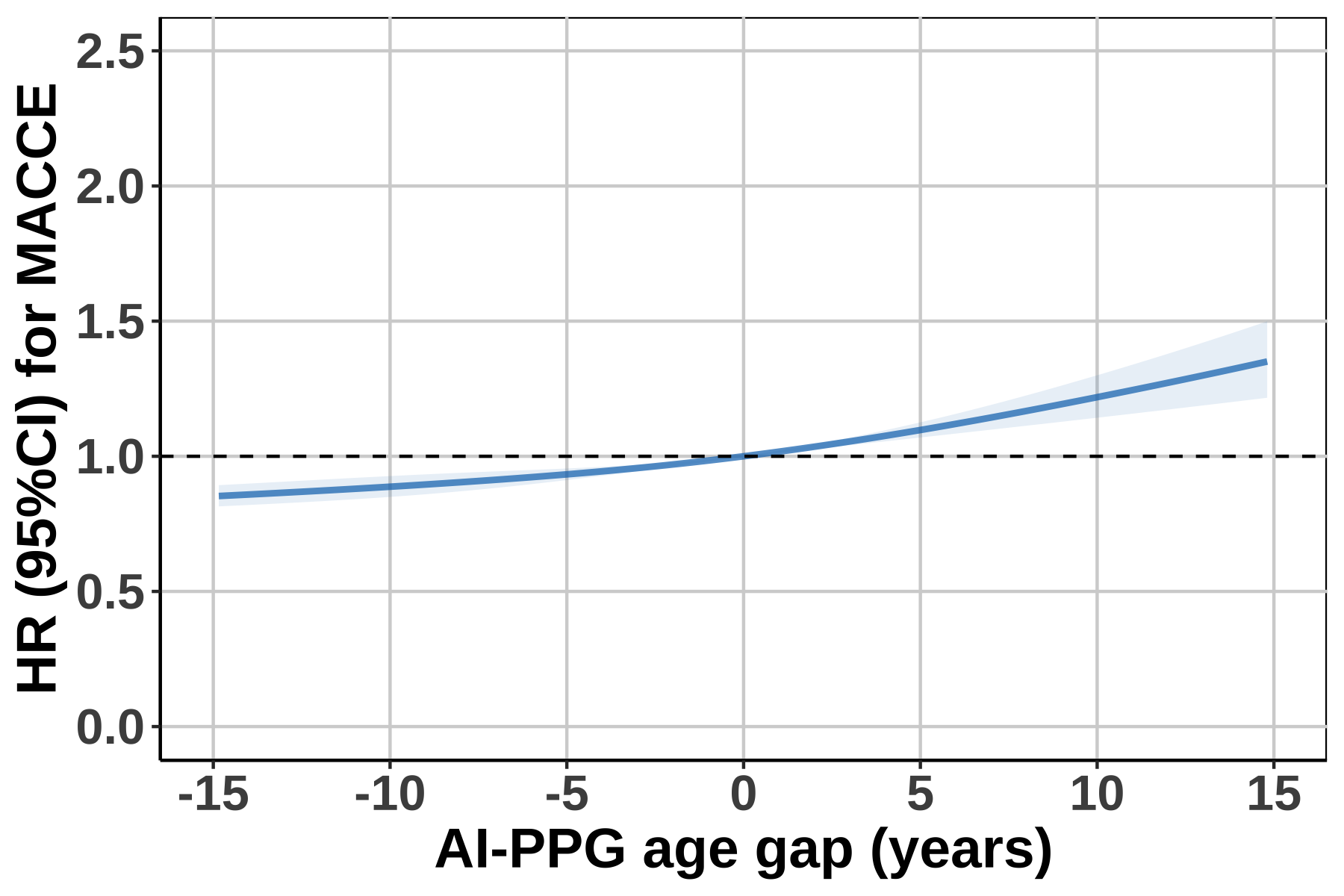}
\end{minipage}%
\hfill
\begin{minipage}{0.24\textwidth}
    \centering
    \includegraphics[width=\linewidth]{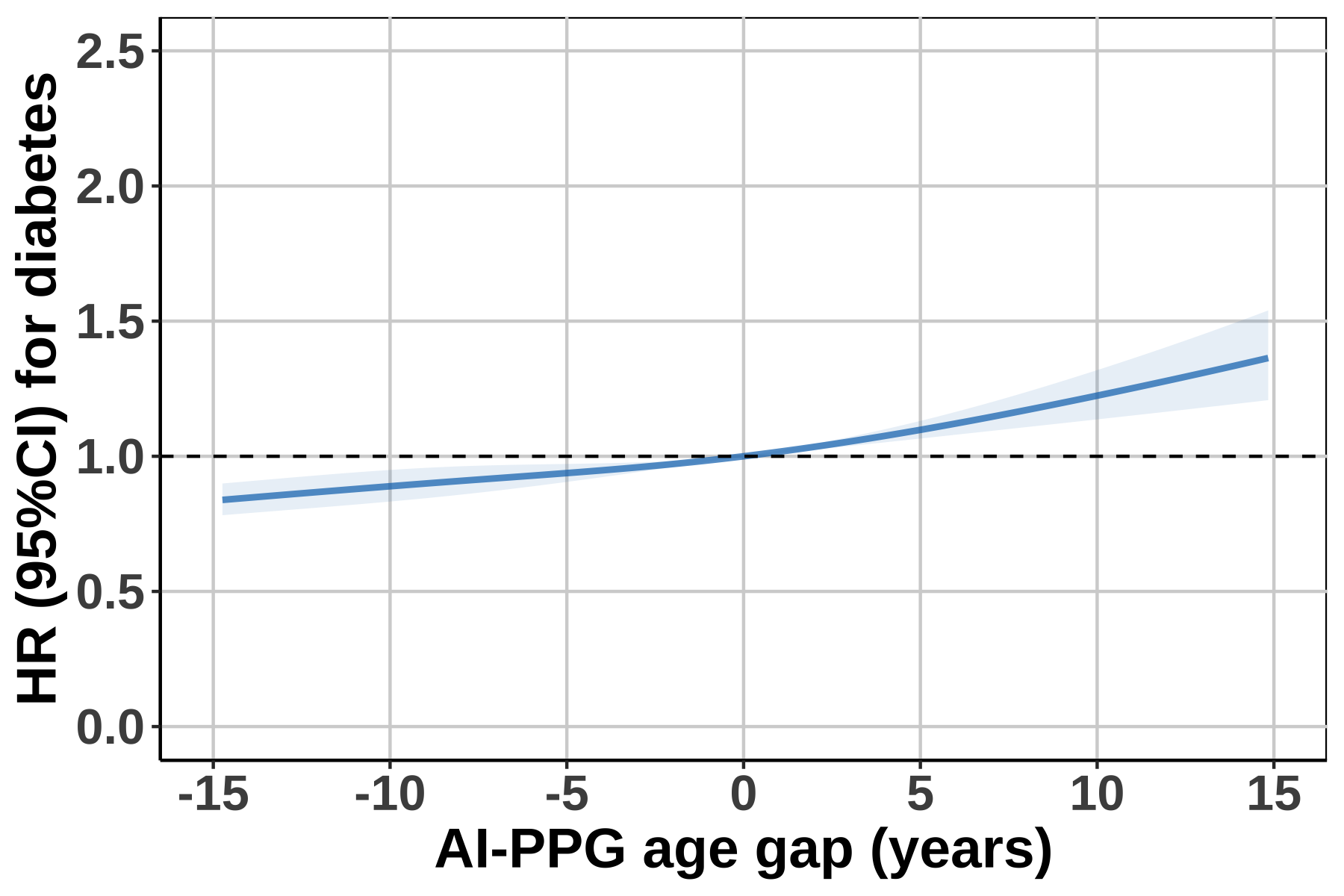}
\end{minipage}%
\hfill
\begin{minipage}{0.24\textwidth}
    \centering
    \includegraphics[width=\linewidth]{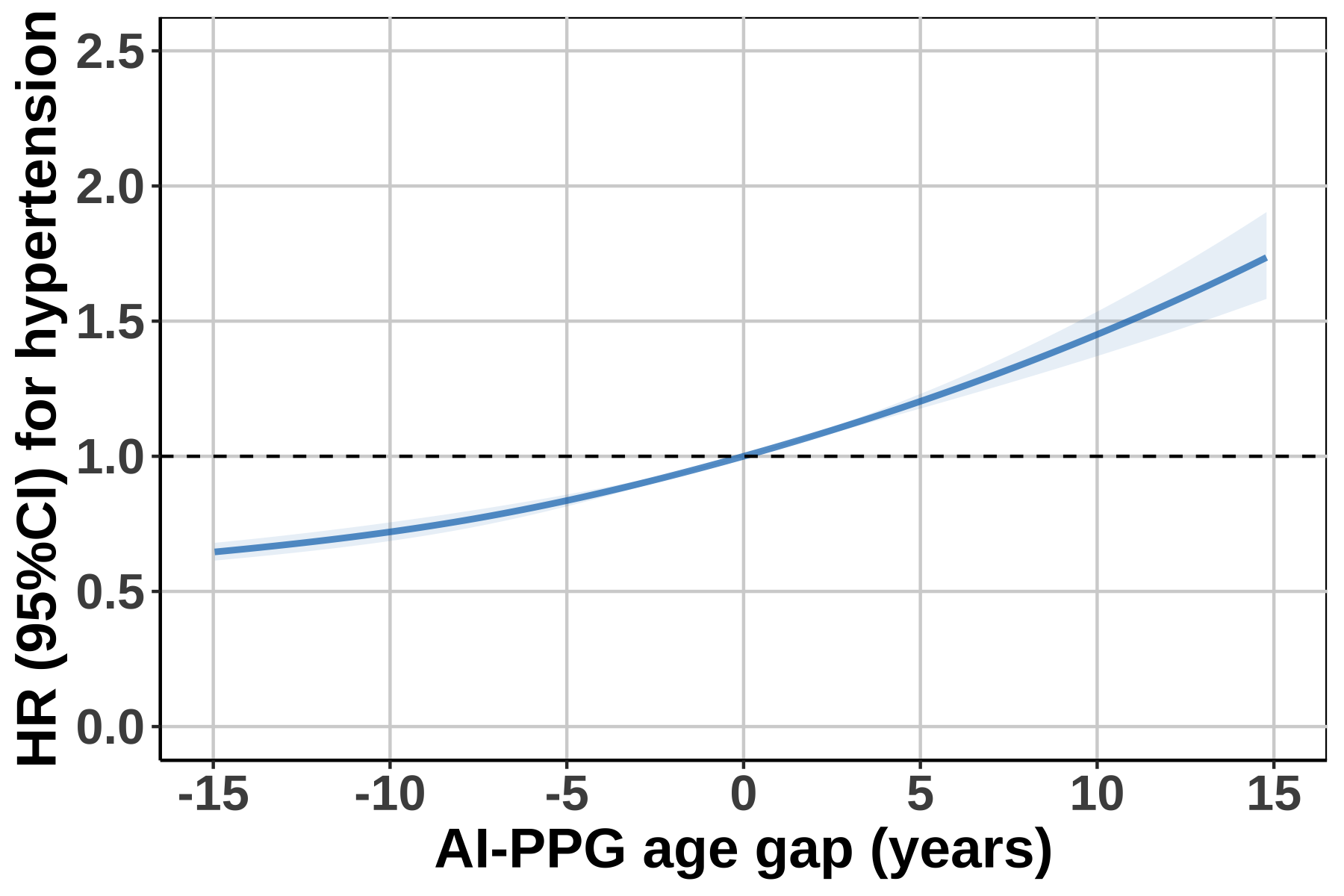}
\end{minipage}%
\hfill
\begin{minipage}{0.24\textwidth}
    \centering
    \includegraphics[width=\linewidth]{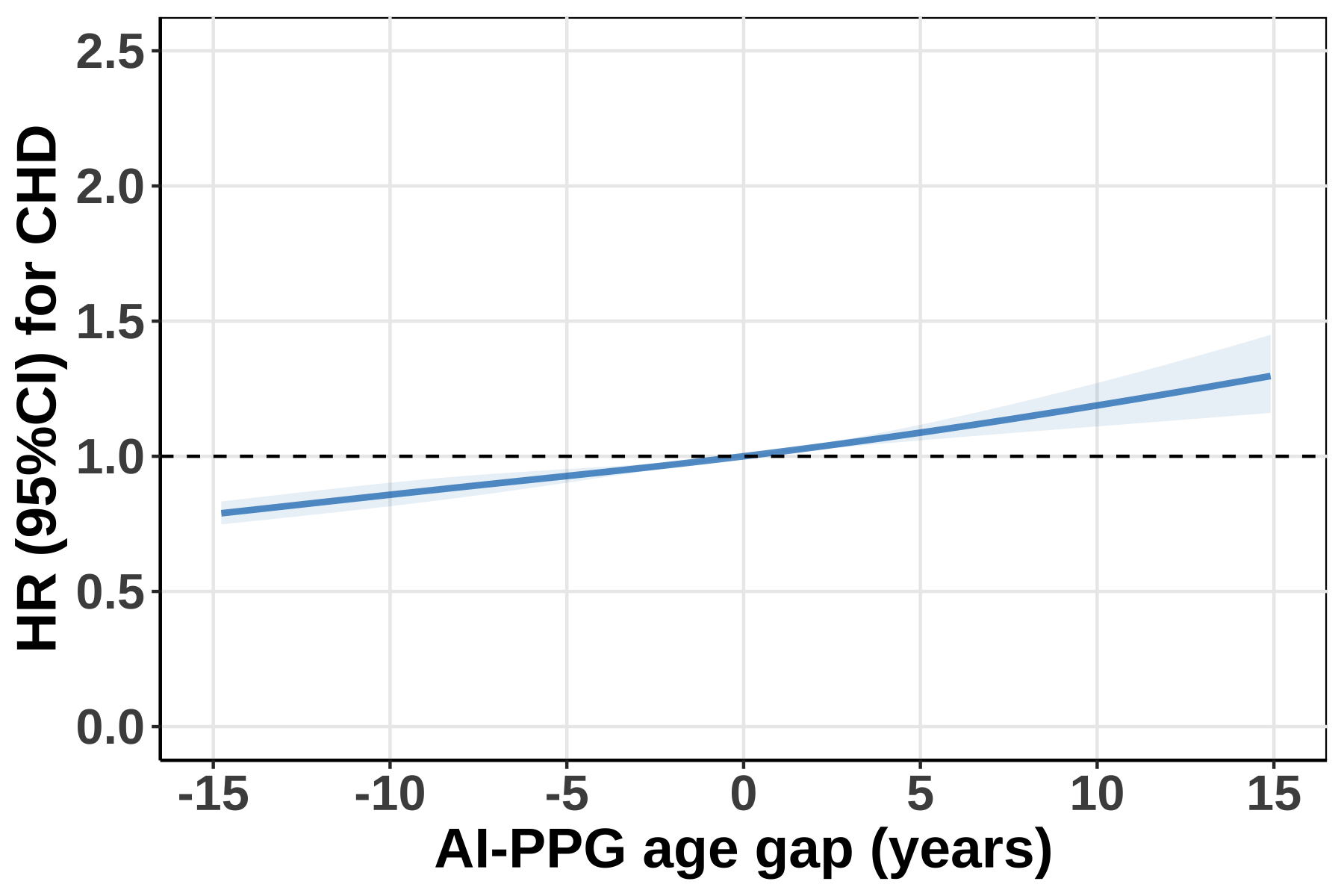}
\end{minipage}%

\vskip 0.3cm

\begin{minipage}{0.24\textwidth}
    \centering
    \includegraphics[width=\linewidth]{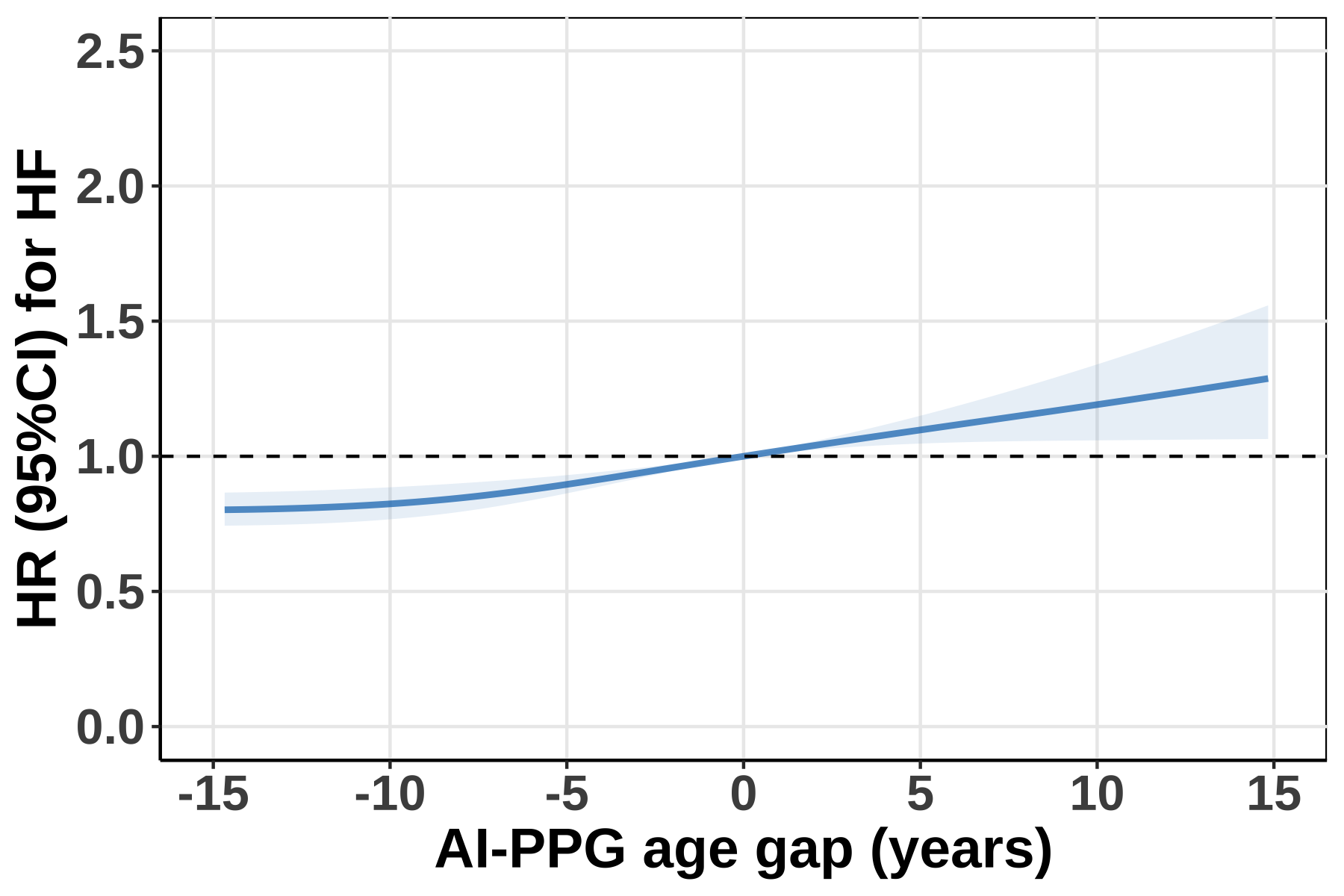}
\end{minipage}%
\hfill
\begin{minipage}{0.24\textwidth}
    \centering
    \includegraphics[width=\linewidth]{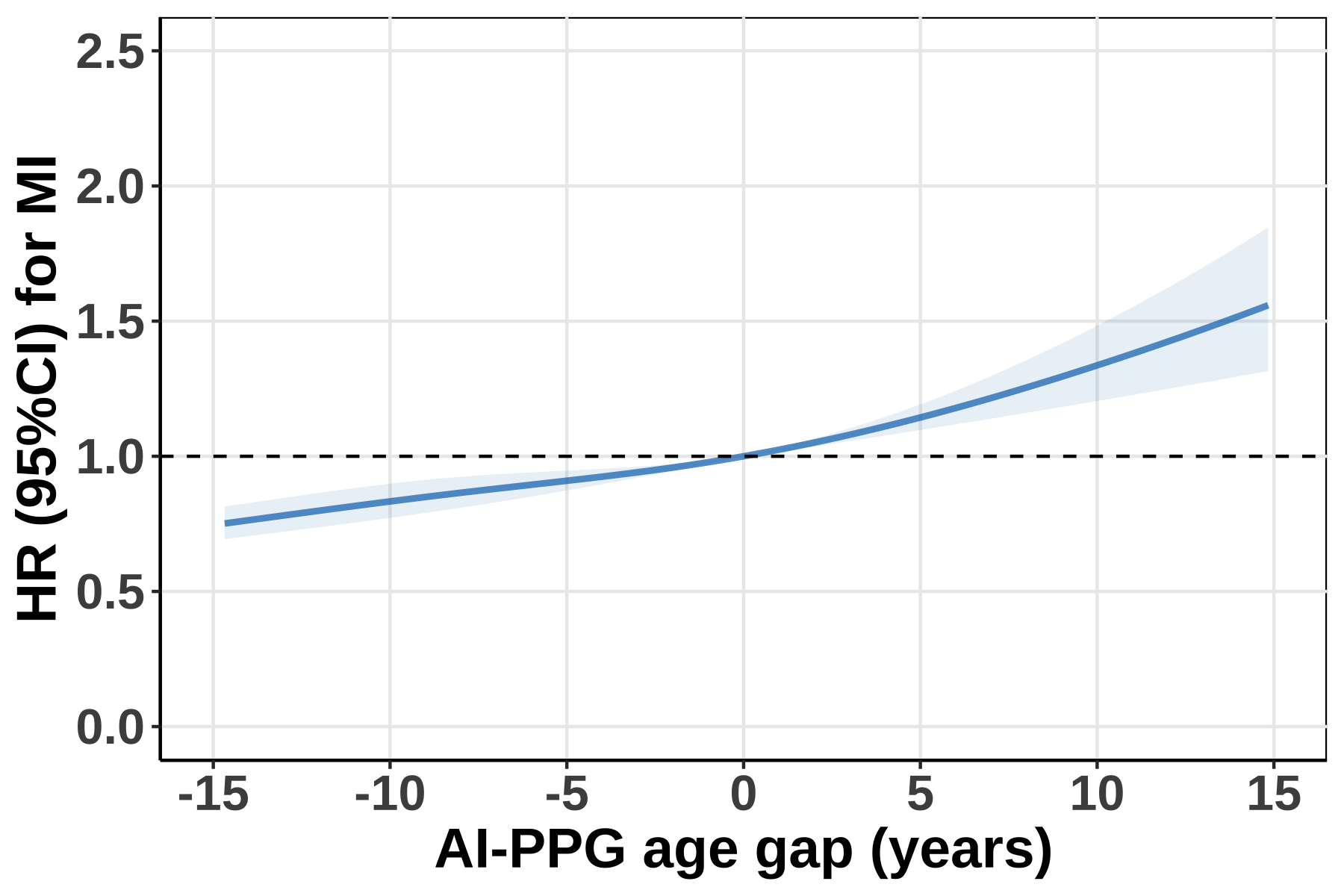}
\end{minipage}%
\hfill
\begin{minipage}{0.24\textwidth}
    \centering
    \includegraphics[width=\linewidth]{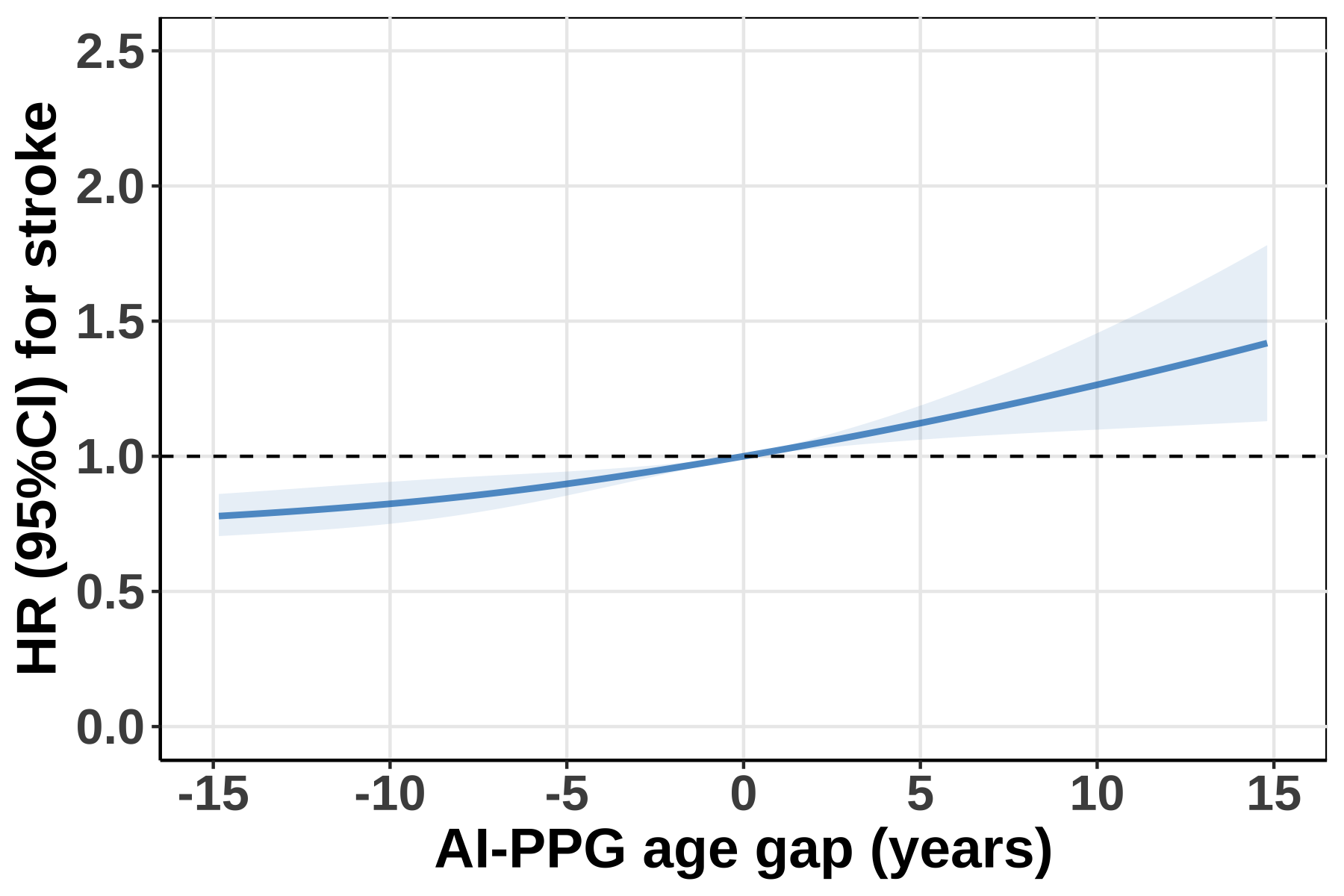}
\end{minipage}%
\hfill
\begin{minipage}{0.24\textwidth}
    \centering
    \includegraphics[width=\linewidth]{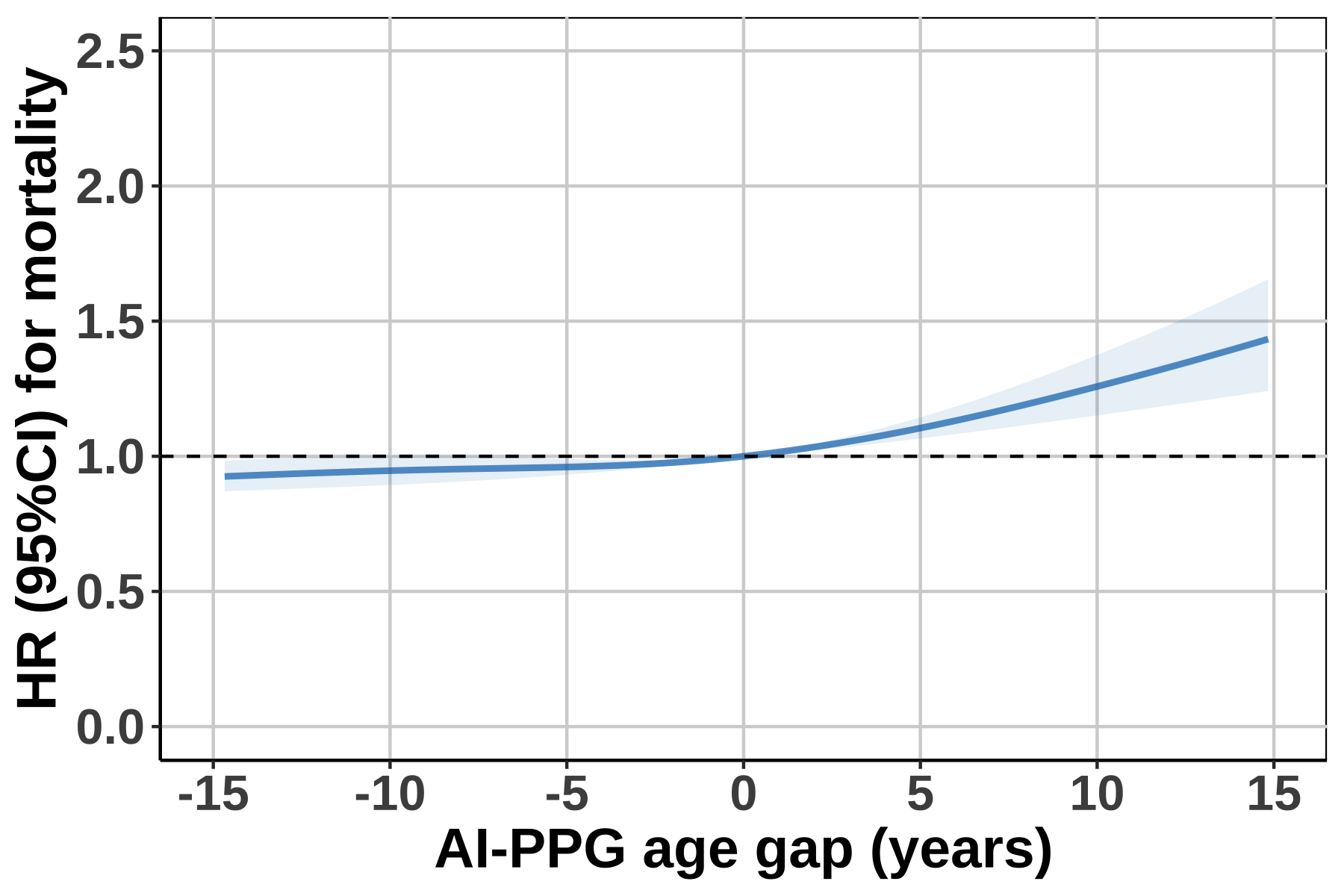}
\end{minipage}%

\caption{Spline curves illustrating the risk of MACCE and cardiometabolic diseases as the AI-PPG age gap increases. MACCE, major adverse cardiovascular and cerebrovascular events.}
\label{Fig: rcs}
\end{figure}

\begin{figure}
\centering
\begin{minipage}{0.24\textwidth}
    \centering
    \includegraphics[width=\linewidth]{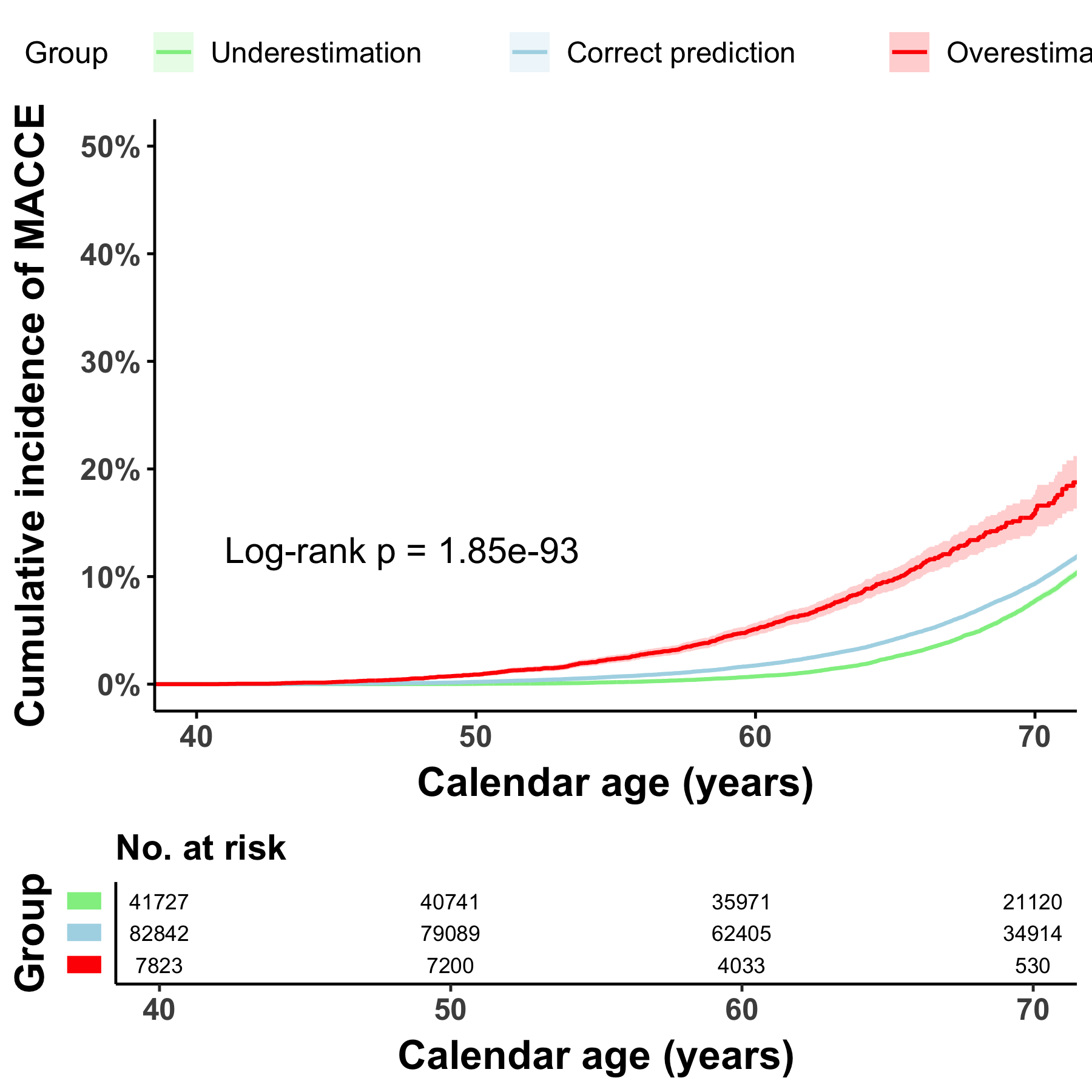}
\end{minipage}%
\hfill
\begin{minipage}{0.24\textwidth}
    \centering
    \includegraphics[width=\linewidth]{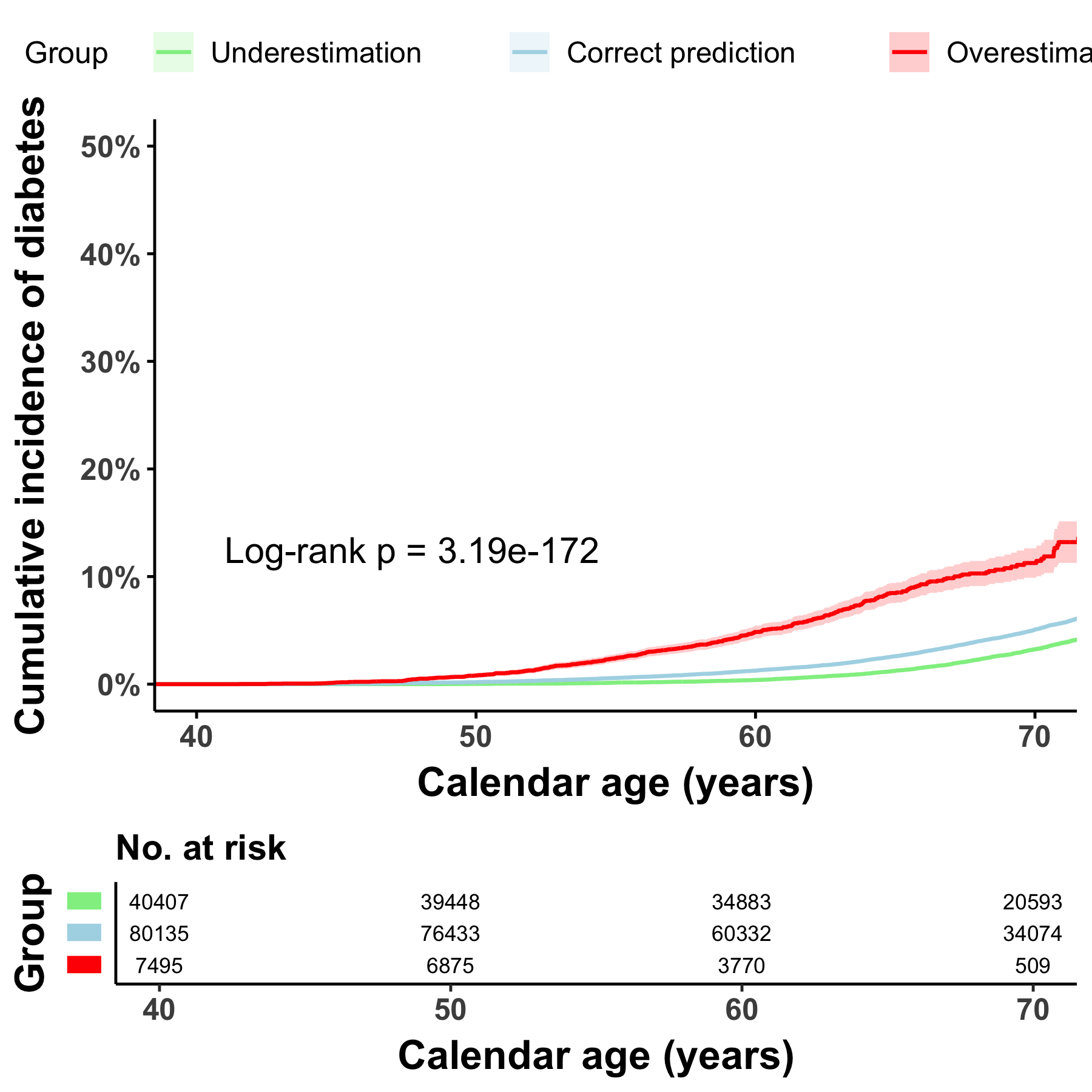}
\end{minipage}%
\hfill
\begin{minipage}{0.24\textwidth}
    \centering
    \includegraphics[width=\linewidth]{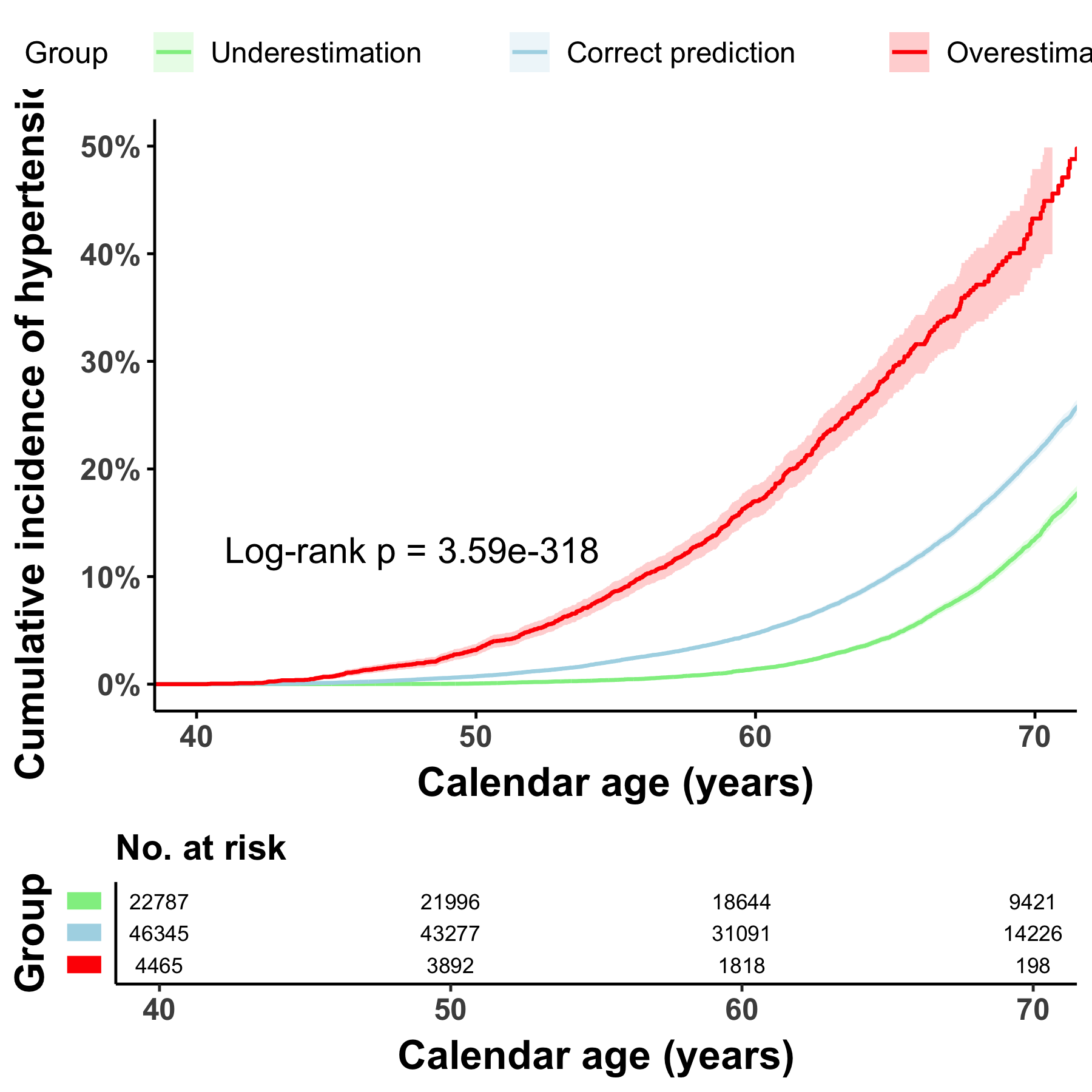}
\end{minipage}%
\hfill
\begin{minipage}{0.24\textwidth}
    \centering
    \includegraphics[width=\linewidth]{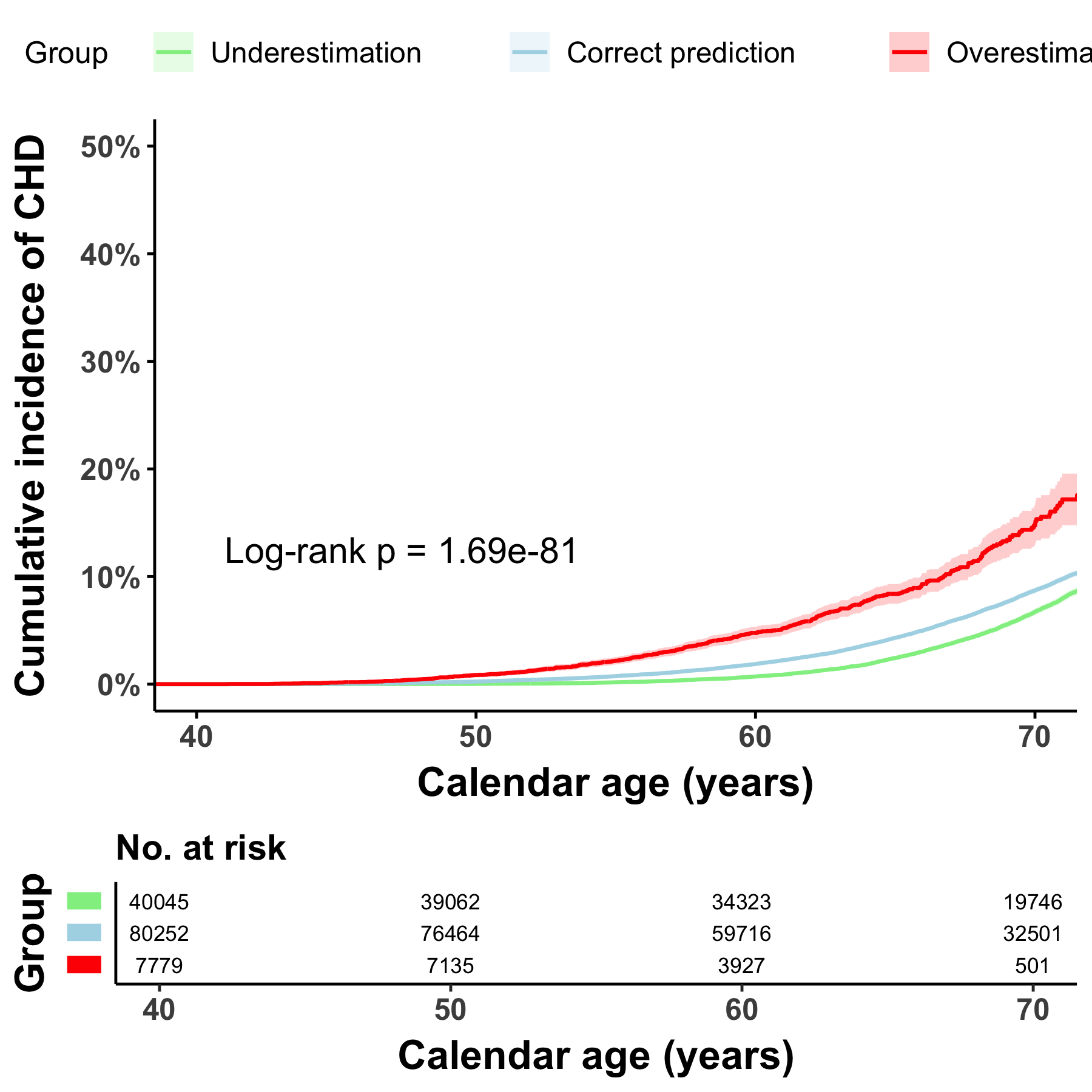}
\end{minipage}%

\vskip 0.3cm 

\begin{minipage}{0.24\textwidth}
    \centering
    \includegraphics[width=\linewidth]{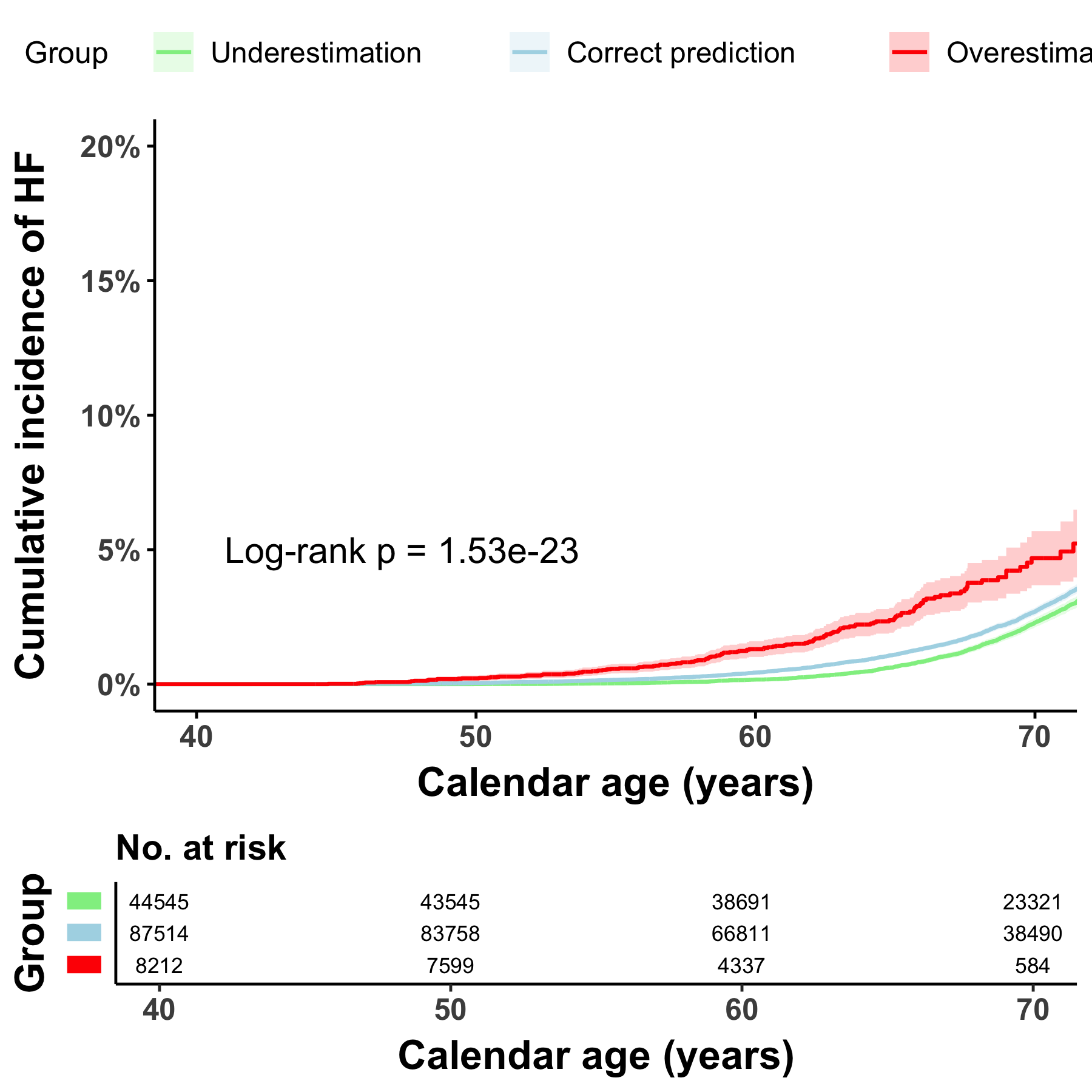}
\end{minipage}%
\hfill
\begin{minipage}{0.24\textwidth}
    \centering
    \includegraphics[width=\linewidth]{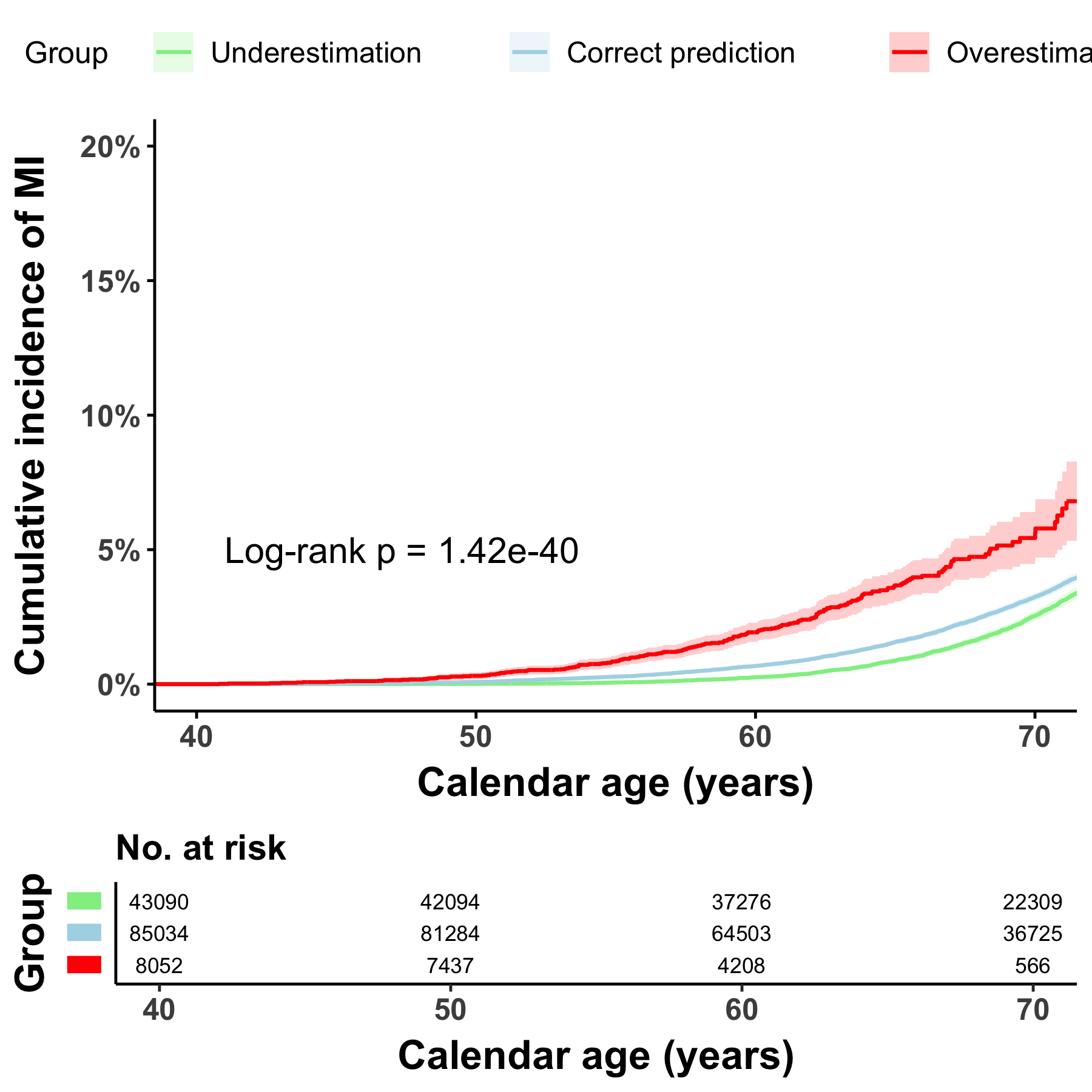}
\end{minipage}%
\hfill
\begin{minipage}{0.24\textwidth}
    \centering
    \includegraphics[width=\linewidth]{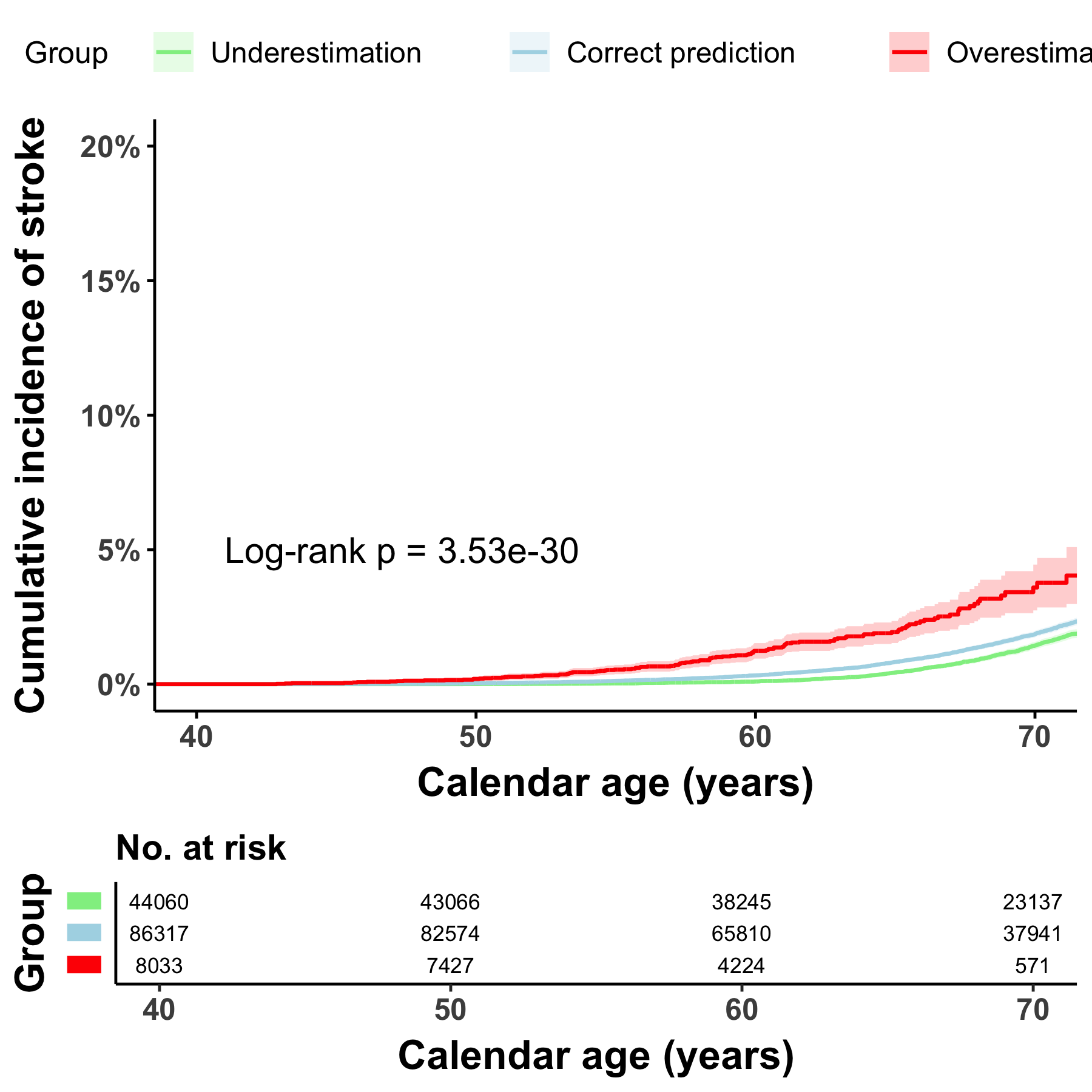}
\end{minipage}%
\hfill
\begin{minipage}{0.24\textwidth}
    \centering
    \includegraphics[width=\linewidth]{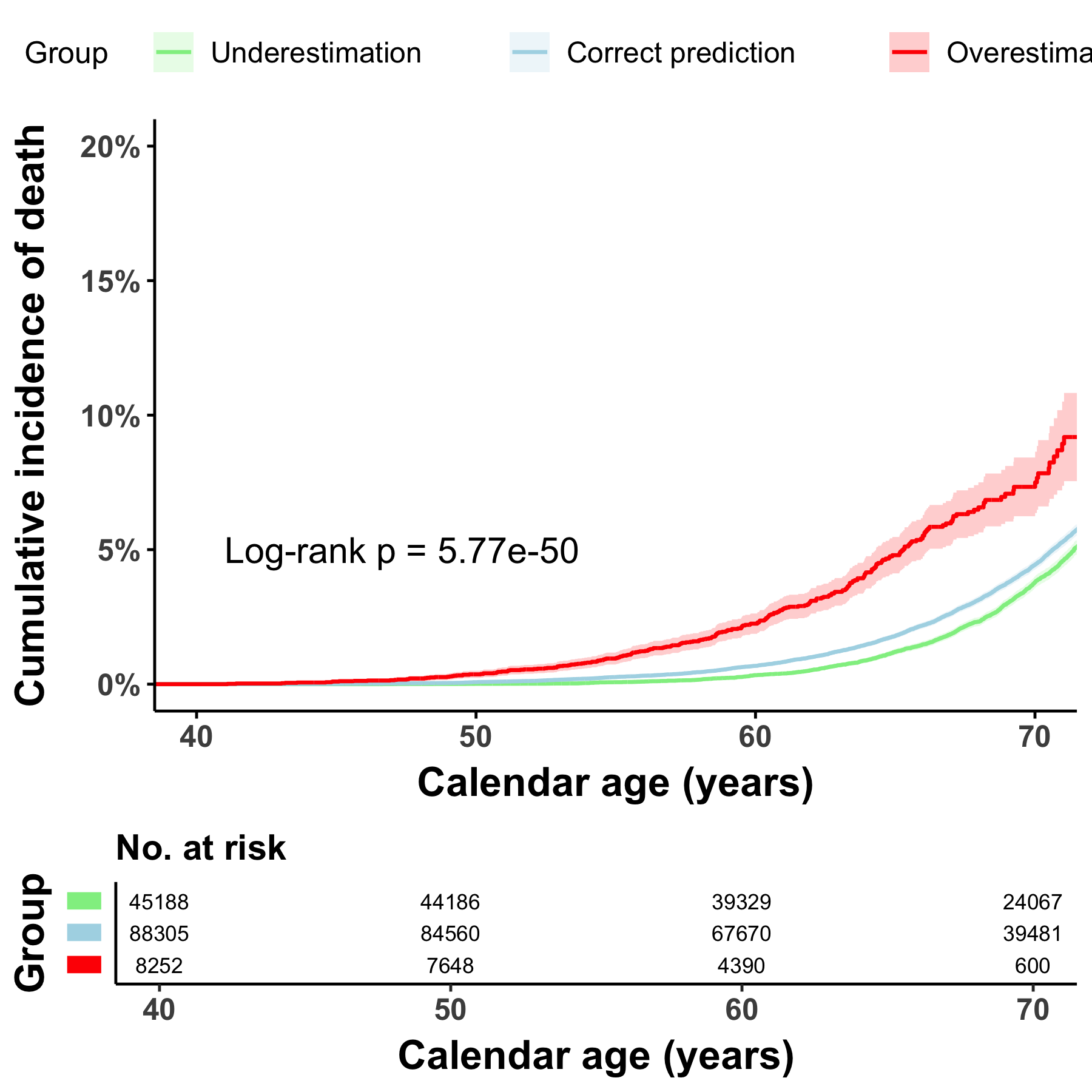}
\end{minipage}%

\caption{KM curves illustrating the cumulative incidence of MACCE and cardiometabolic diseases across three groups: overestimation, correct prediction, and underestimation. KM, Kaplan-Meier; MACCE, major adverse cardiovascular and cerebrovascular events.}
\label{Fig: km}
\end{figure}

\subsection*{AI-PPG age gap classification over time based on serial PPG data is associated with future risk of major adverse cardiovascular and cerebrovascular events}
The results based on serial PPG measurements at two time points are shown in Fig. \ref{Fig: serialPPG}. Group G4 (both measurements classified as correct prediction or underestimation) served as the reference. The KM curves and corresponding HRs for future MACCE are presented for Group G1 (both measurements classified as overestimation), Group G2 (first measurement classified as underestimation or correct prediction, second as overestimation), and Group G3 (first measurement classified as overestimation, second as underestimation or correct prediction). \textcolor{blue}{The HRs were 5.53 (95\% CI: 2.73–11.23, p = 1.50$\times$10$^{-6}$) for Group G1, 3.63 (95\% CI: 1.86–7.06, p = 2.67$\times$10$^{-4}$) for Group G2, and 3.02 (95\% CI: 2.21–4.13, p = 1.23$\times$10$^{-9}$) for Group G3.} The log-rank test p-values for comparisons of G1, G2, and G3 with the reference group G4 were all < 0.001, indicating significant differences. Notably, the HRs were consistently higher than 1 and and showed an increasing trend from Group G3 to Group G1, suggesting that changes in AI-PPG age gap classification over time may provide additional value beyond single-time-point assessments.

\begin{figure}
\centering
\begin{minipage}{0.33\textwidth}
    \centering
    \includegraphics[width=\linewidth]{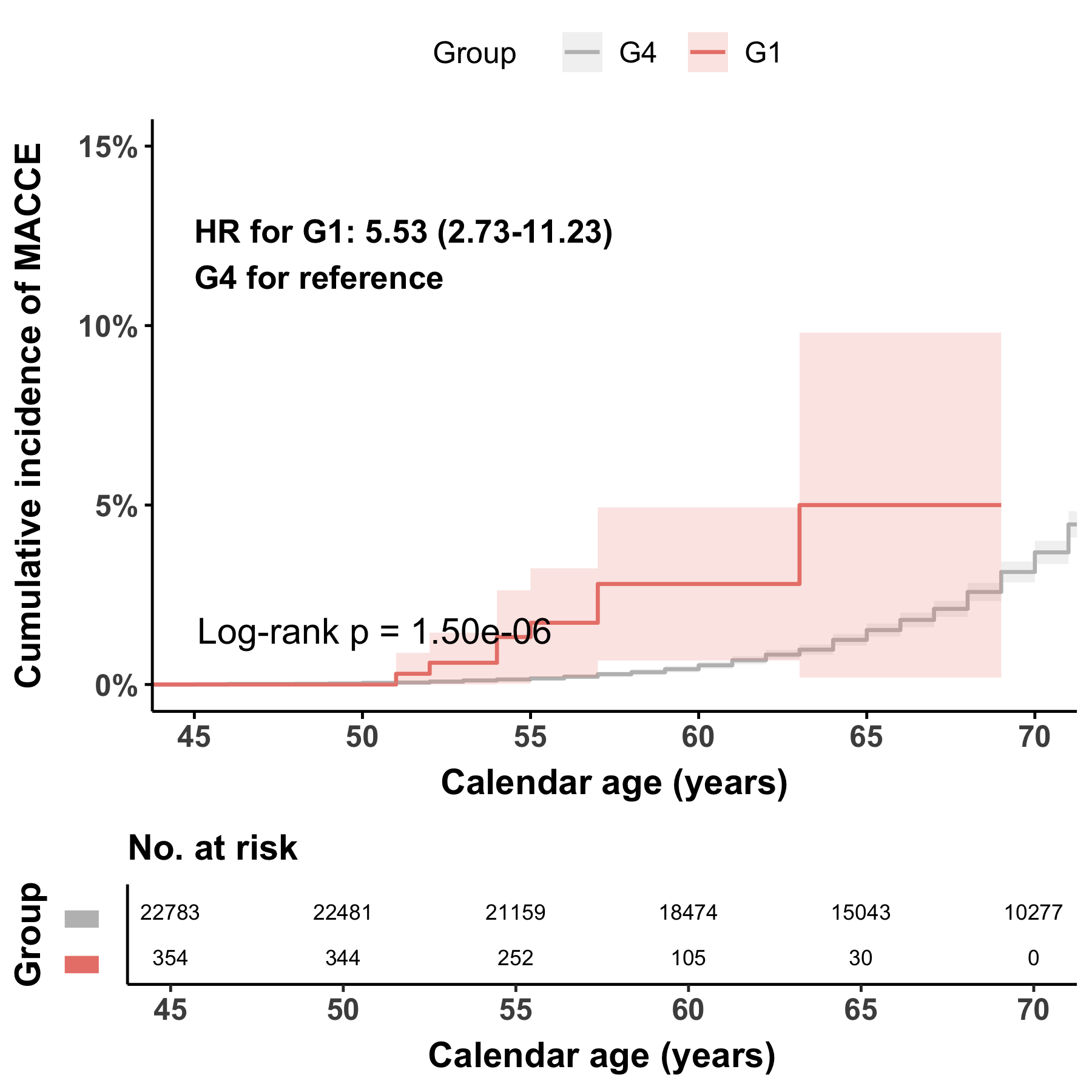}
\end{minipage}%
\hfill
\begin{minipage}{0.33\textwidth}
    \centering
    \includegraphics[width=\linewidth]{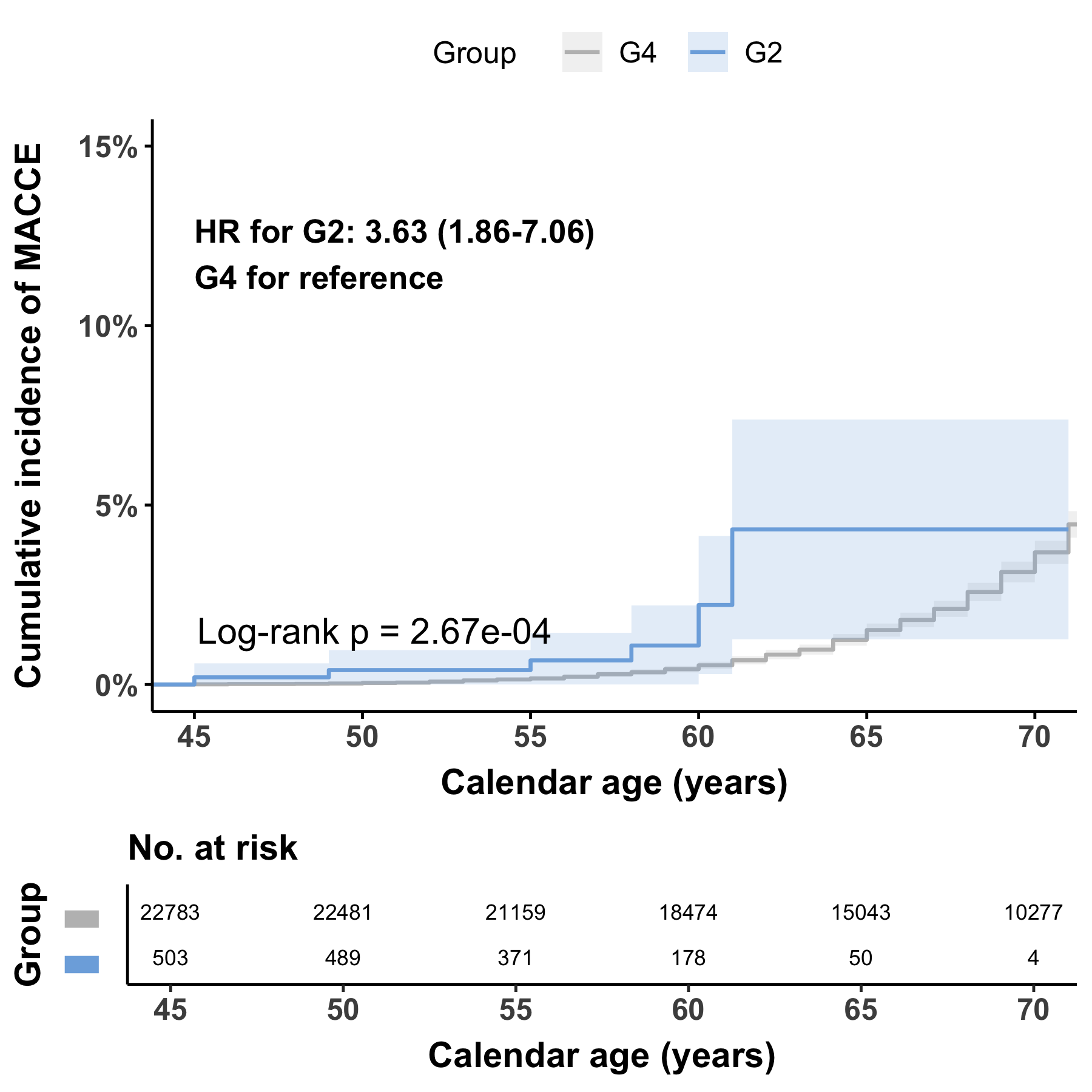}
\end{minipage}%
\hfill
\begin{minipage}{0.33\textwidth}
    \centering
    \includegraphics[width=\linewidth]{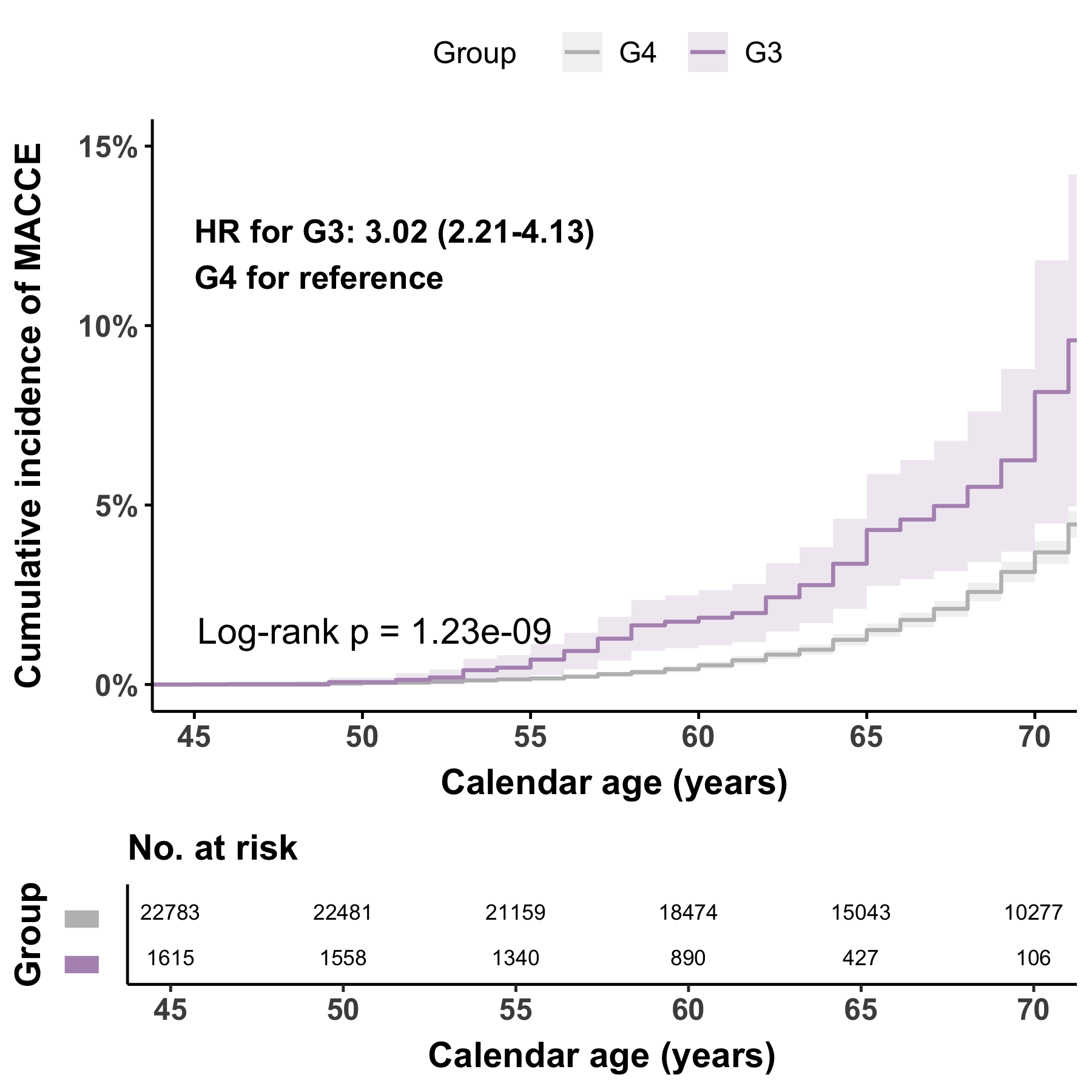}
\end{minipage}%
\caption{KM curves for future MACCE based on AI-PPG age gap classifications from serial PPG measurements at two time points. Group G4 (reference) represents individuals with both measurements classified as correct prediction or underestimation. Group G1 includes individuals with both measurements classified as overestimation, Group G2 includes those with the first measurement classified as underestimation or correct prediction and the second as overestimation, and Group G3 includes those with the first measurement classified as overestimation and the second as underestimation or correct prediction. The KM curves compare G1, G2, and G3 to the reference group (G4), with corresponding HRs indicated on each plot. KM, Kaplan-Meier; MACCE: major adverse cardiovascular and cerebrovascular events; HR: hazard ratio.}
\label{Fig: serialPPG}
\end{figure}

\subsection*{Overestimated AI-PPG age correlates with increased in-hospital mortality in critical ill patients}
In the external validation using the MIMIC hospital mortality set, we evaluated the association between vascular age gap and in-hospital mortality. AI-PPG age gap was analyzed both as a continuous variable and a categorical variable (Table \ref{Tab: MIMIC HRs}). For the categorical analysis, underestimation and overestimation were defined using a threshold of 15 years, based on the standard deviation of the overall prediction error.

After adjusting for age, sex, and ethnic background, AI-PPG age gap as a continuous variable was significantly associated with in-hospital mortality, with an odds ratio (OR) of 1.02 per one-year increase (p = 0.01). When treated as a categorical variable, underestimation showed no significant difference compared to the correct prediction group (OR: 0.91, p = 0.49). In contrast, overestimation was associated with a significantly higher risk of in-hospital mortality (OR: 1.86, \textcolor{blue}{p = 4.04$\times$10$^{-4}$}). Further adjustments for smoking status, hypertension, diabetes, dyslipidemia, and CKD revealed that each additional year of AI-PPG age gap remained marginally associated with increased in-hospital mortality (OR: 1.01, p = 0.06). The underestimation group continued to show no significant difference from the correct prediction group (OR: 0.95, p = 0.71), whereas overestimation remained significantly associated with a higher risk of in-hospital mortality (OR: 1.74, p = 0.01). These findings suggest that a greater AI-PPG age gap, particularly in cases of overestimation, is linked to an elevated risk of in-hospital mortality.

\begin{table}[]
\resizebox{\linewidth}{!}{
\begin{threeparttable}
\caption{Association between AI-PPG age gap and in-hospital mortality in the MIMIC hospital mortality set. AI-PPG age gap is analyzed as a continuous variable, while underestimation and overestimation are categorized relative to the correct prediction group. OR, odds ratio.}
\label{Tab: MIMIC HRs}
\centering
\renewcommand{\arraystretch}{1.2} 
\begin{tabular}{ccccccc}
\hline
                      & \multicolumn{2}{c}{AI-PPG age gap} & \multicolumn{2}{c}{Underestimation} & \multicolumn{2}{c}{Overestimation} \\ \cline{2-7} 
                      & OR (95\% CI)          & p-value      & OR (95\% CI)          & p-value     & OR (95\% CI)     & p-value         \\ \hline
\multicolumn{7}{l}{Model 1, adjusted for age, sex, and ethnicity}                                                                        \\
In-hospital mortality & 1.02 (1.00-1.03)      & 0.01         & 0.91 (0.69-1.19)      & 0.49        & 1.86 (1.24-2.79) & 4.04 $\times$ 10$^{-4}$ \\ \hline
\multicolumn{7}{l}{Model 2, adjusted for age, sex, ethnicity, smoking status, hypertension, diabetes, dyslipidemia, and CKD}             \\
In-hospital mortality & 1.01 (1.00-1.03)      & 0.06         & 0.95 (0.72-1.25)      & 0.71        & 1.74 (1.15-2.63) & 0.01            \\ \hline
\end{tabular}
\begin{tablenotes}
\footnotesize
\item  \textcolor{blue}{Results in the table were obtained using Cox proportional hazards regression.}
\end{tablenotes}
\end{threeparttable}}
\end{table}

\subsection*{AI-PPG age model demonstrates consistent focus on key PPG features across age groups}
Figure \ref{Fig: saliency map} presents the average PPG waveforms and corresponding saliency maps for three distinct age groups (40, 55, and 70 years old). In the saliency maps, regions with higher brightness indicate areas that attract more attention from the model and are therefore more influential in the prediction process. The saliency maps show that, across all age groups, the model consistently focuses on the region surrounding the diastolic peak, with secondary attention directed towards the systolic peak. This suggests that the model relies on these two key features in the PPG waveform for age prediction. Correspondingly, the average PPG waveforms exhibit age-related changes, particularly in the diastolic and systolic peaks. As age increases, the distance between the two peaks narrows, and the prominence of the diastolic peak gradually diminishes. These observed trends in the PPG waveforms closely match the areas of attention highlighted in the saliency maps, which is consistent with findings in existing literature \cite{yousef2012analysis, lin2023age}. This alignment further supports the model’s ability to identify features related to vascular aging in the PPG signals, contributing to accurate age estimation.

\begin{figure}
    \centering
    \includegraphics[width=0.96\linewidth]{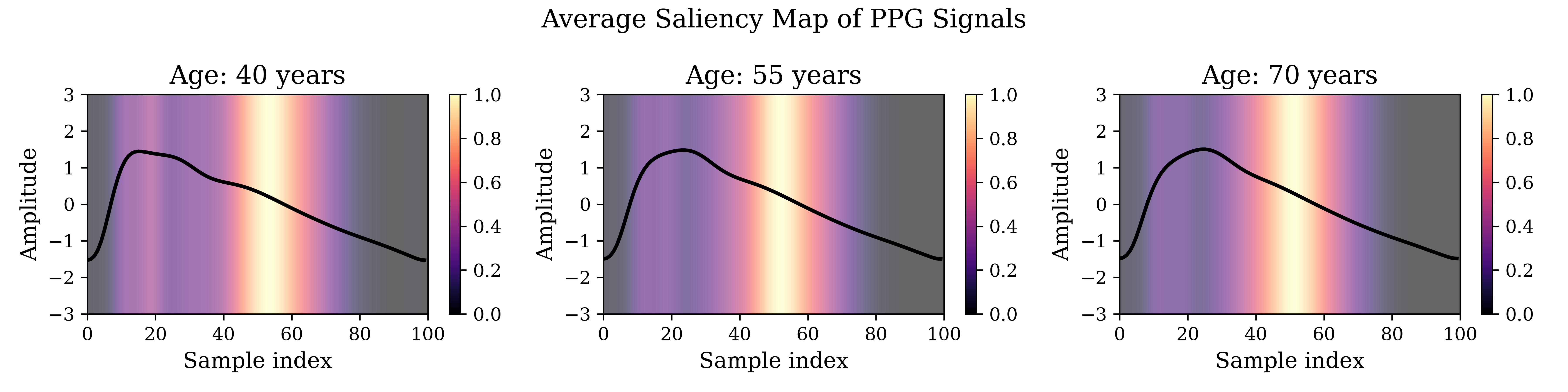}
    \caption{Averaged PPG signals and corresponding saliency maps across three age groups (40, 55, and 70 years old). Each column represents the averaged PPG signal and its associated saliency map for the respective age group.}
    \label{Fig: saliency map}
\end{figure}

\section*{Discussion}
In this study, we developed a deep learning-based framework to estimate AI-PPG age for streamlined cardiovascular health assessment. The model was trained and evaluated on the UKB dataset and externally validated using the MIMIC subset of the PulseDB dataset. Our results demonstrate that the AI-PPG age gap, defined as the difference between AI-PPG age and chronological age, serves as an independent risk factor for the primary outcome MACCE and secondary outcomes, including mortality, HF, MI, stroke, CHD, hypertension, and diabetes in the UKB cohort. External validation further confirmed its association with in-hospital mortality in the MIMIC subset, underscoring its potential clinical utility in both general and critically ill populations.

We also conducted an analysis of serial PPG measurements within the UKB cohort. Participants were categorized into four groups based on the AI-PPG age gap classifications at two time points: G1 (overestimation in both measurements), G2 (not overestimation in the first measurement but overestimation in the second), G3 (overestimation in the first measurement but not in the second), and G4 (not overestimation in either measurement). Compared to G4, participants in the other groups exhibited significantly higher risks for future MACCE events: \textcolor{blue}{G1 (HR 5.53, 95\% CI: 2.73–11.23, p = 1.50$\times$10$^{-6}$), G2 (HR 3.63, 95\% CI: 1.86–7.06, p = 2.67$\times$10$^{-4}$), and G3 (HR 3.02, 95\% CI: 2.21–4.13, p = 1.23$\times$10$^{-9}$).} The HR showed a progressively increasing trend from G3 to G1. This pattern may reflect the cumulative effect of vascular aging. G1 participants consistently displayed advanced vascular aging across both measurements, signifying prolonged exposure to poor cardiovascular health and the highest overall risk. In G2, the transition to overestimation during the second measurement suggests an accelerated progression of vascular aging, which reflects a recent deterioration in cardiovascular health. Consequently, their risk is elevated but remains lower than that of G1. For G3, although the second measurement indicates improvement in cardiovascular health, the long-term effects of previous excessive vascular aging may persist, leading to a moderately increased risk compared to G4, but still lower than G2. These results demonstrate the potential of addictive value of serial PPG data. These findings highlight the additive value of serial PPG measurements in capturing dynamic changes in vascular aging and associated cardiovascular risk.

To address the common challenge of data imbalance in regression tasks \cite{yang2021delving}, particularly the age imbalance in the UKB dataset, we propose a loss function, Dist Loss. This method aligns the distribution of model predictions with the ground truth label distribution, effectively mitigating the risk of over-predicting samples from underrepresented regions into areas with higher label density. Our approach provides a straightforward and effective solution for handling imbalanced data distributions, offering a practical and reliable method for regression tasks under such conditions.

In tackling the interpretability challenges often associated with deep learning models, this study utilizes saliency map visualizations to identify areas of focus within the input signals. The results reveal that the model predominantly concentrates on the region around the diastolic peak of the PPG waveform when predicting AI-PPG age, with secondary emphasis on the systolic peak. These findings are consistent with observed variations in the average PPG waveforms across different age groups. This approach provides a useful framework for understanding and explaining the model's prediction behavior, supporting conclusions from existing research \cite{yousef2012analysis, lin2023age}.

CVDs remain the leading global health threat, and empowering individuals to understand and manage their cardiovascular health is essential to reducing the associated societal burden \cite{vogel2021lancet, khan2023novel}. The rapid proliferation of wearable devices, particularly those equipped with embedded PPG sensors, offers a promising avenue for non-invasive, convenient, and continuous hemodynamic monitoring \cite{wahab2025systematic, williams2023wearable}. In this context, we introduce AI-PPG age, an AI-derived biomarker representing biological age that serves as an intuitive and interpretable proxy for vascular health, promoting individual awareness and self-monitoring. Importantly, AI-PPG age is derived exclusively from raw PPG signals, enabling seamless integration into wearable platforms such as smartwatches. This approach offers two key clinical benefits: first, it supports large-scale cardiovascular screening and risk stratification through wearable devices; second, it facilitates targeted clinical follow-up and personalized management for individuals identified as high risk. By enabling early detection of vascular health changes and guiding timely interventions, this tiered strategy holds significant potential to improve cardiovascular outcomes.

This study has several limitations. First, the results demonstrated a moderate correlation between AI-PPG age and calender age, which may be attributed to the limited age-related information captured in the relatively short PPG recordings. While this does not compromise the utility of AI-PPG age in cardiovascular risk stratification, improving its numerical alignment with calender age could enhance interpretability in practical applications. One potential approach is to apply post hoc correction strategies, such as the bias adjustment methods proposed in previous studies \cite{beheshti2019bias, le2022using, le2018nonlinear}. Second, the waveform of PPG signals is not only affected by vascular aging but also by other physiological factors such as blood pressure and the use of vasoactive medications. In this study, to ensure ease of use and convenience, we developed the AI-PPG age prediction model using only raw PPG signals without incorporating additional covariates. While this design improves applicability, it may limit the model's ability to comprehensively capture cardiovascular health. Future work may explore combining AI-PPG age with other cardiovascular-related indicators at the application stage to further improve assessment performance. Third, following previous studies \cite{al2025advanced, cho2025artificial, lima2021deep}, we used the standard deviation of the AI-PPG age gap as a threshold for stratifying individuals. However, this data-driven threshold may vary across populations, raising concerns about its generalizability. While our findings in the UKB cohort support a 9-year threshold for the general population, further validation is needed before applying this cutoff in other demographic or clinical settings. Finally, AI-PPG age is intended to serve as an auxiliary indicator of cardiovascular risk rather than a standalone diagnostic tool. Its use in real-world settings should be integrated with other clinical information to avoid over-intervention due to outliers or underestimation of risk due to occasional prediction errors.

Despite these limitations, our study demonstrates that AI-PPG age is a scalable and non-invasive digital biomarker with potential for broad implementation in wearable devices. By supporting continuous cardiovascular health monitoring and enabling early risk stratification at the population level, AI-PPG age offers a promising approach to enhance preventive cardiology and inform personalized clinical decision-making.

\section*{Data Availability}
This study used data from the UK Biobank. \textcolor{blue}{Access to UK Biobank data is not publicly available and requires an application.} Data are provided to approved researchers worldwide (\url{https://www.ukbiobank.ac.uk/}). 
The PulseDB dataset is publicly available at \url{https://github.com/pulselabteam/PulseDB}. 
Access to the MIMIC-III Clinical Database requires completion of the Collaborative Institutional Training Initiative course and approval from PhysioNet (\url{https://physionet.org/content/mimiciii/1.4}).

\section*{Code Availability}
\textcolor{blue}{The implementation of Dist Loss is available in the GitHub repository at \url{https://github.com/Ngk03/Dist-Loss} \cite{nie2025distloss}. The model configuration and pretrained weights can be found at \url{https://huggingface.co/Ngks03/PPG-VascularAge} \cite{nie2025ppgvascularage}.}

\section*{Competing interests}
The authors declare no competing interests.

\section*{Acknowledgement}
This work was supported by the Beijing Natural Science Foundation (QY23040), the National Natural Science Foundation of China (62102008), and the Research Project of Peking University in the State Key Laboratory of Vascular Homeostasis and Remodeling (2025-SKLVHR-YCTS-02).

\section*{Author Contributions} 
G.N. performed the primary data processing and statistical analyses, and drafted the initial version of the manuscript. Q.Z. contributed to the design of the validation framework, carried out additional data analyses, and critically revised the manuscript. G.T. assisted with algorithm implementation and execution of experiments. Y.L. contributed to data aggregation and baseline preparation. S.H. conceived and supervised the overall study, provided critical revisions to the manuscript, and secured access to data and computational resources.

\newpage


\bibliographystyle{naturemag} 
\bibliography{reference}

\bigskip

\end{document}